\documentclass[twocolumn,tighten,times]{aastex631}
\usepackage{xspace}
\usepackage{amsmath}
\def\hi{\textsc{Hi}\xspace}
\newcommand{\sofia}{\texttt{SoFiA}\xspace}
\newcommand{\sofiaii}{\texttt{SoFiA2}\xspace}
\usepackage{color}
\usepackage{cases}
\usepackage{enumitem}
\usepackage{threeparttable}

\newcommand{\kms}{\,km\,s$^{-1}$}

\newcommand{\Mpc}{\ensuremath{\,{\rm Mpc}}}
\newcommand{\K}{\ensuremath{\, {\rm K}}}

\newcommand{\Jy}{\ensuremath{\,{\rm Jy}}}
\newcommand{\mJy}{\ensuremath{\,{\rm mJy}}}

\renewcommand{\deg}{\ensuremath{\,{\rm deg}}}

\newcommand{\reffg}[1]{Fig.~\ref{#1}}
\newcommand{\reftb}[1]{Table~\ref{#1}}
\newcommand{\refeq}[1]{Eq.~(\ref{#1})}
\newcommand{\refsc}[1]{Sec.~\ref{#1}}

\defcitealias{Li2023}{L23}

\begin{document}

\title{FATHOMER survey: III. Preliminary \textsc{Hi} galaxy identification results}

\author[0009-0004-8919-7088]{Shuanghao Shu}
\affiliation{State Key Laboratory of Radio Astronomy and Technology, National Astronomical Observatories, CAS, A20 Datun Road, Chaoyang District, Beijing, 100101, P. R. China\\}
\affiliation{School of Astronomy and Space Science, University of Chinese Academy of Sciences, Beijing 100049, China\\}

\author[0000-0003-1962-2013]{Yichao Li}
\affiliation{Key Laboratory of Cosmology and Astrophysics (Liaoning)
\& College of Sciences, Northeastern University, Shenyang 110819, China\\}

\author[0009-0006-2521-025X]{Wenxiu Yang}
\affiliation{State Key Laboratory of Radio Astronomy and Technology, National Astronomical Observatories, CAS, A20 Datun Road, Chaoyang District, Beijing, 100101, P. R. China\\}
\affiliation{School of Astronomy and Space Science, University of Chinese Academy of Sciences, Beijing 100049, China\\}

\author[0009-0008-7631-7991]{Jiaxin Wang}
\affiliation{Key Laboratory of Cosmology and Astrophysics (Liaoning)
\& College of Sciences, Northeastern University, Shenyang 110819, China\\}

\author[0000-0002-3108-5591]{Wenkai Hu}
\affiliation{State Key Laboratory of Radio Astronomy and Technology, National Astronomical Observatories, CAS, A20 Datun Road, Chaoyang District, Beijing, 100101, P. R. China\\}
\affiliation{Department of Physics and Astronomy, University of the Western Cape, Robert Sobukwe Road, Bellville, 7535, South Africa\\}

\author[0000-0001-8075-0909]{Furen Deng}
\affiliation{State Key Laboratory of Radio Astronomy and Technology, National Astronomical Observatories, CAS, A20 Datun Road, Chaoyang District, Beijing, 100101, P. R. China\\}
\affiliation{School of Astronomy and Space Science, University of Chinese Academy of Sciences, Beijing 100049, China\\}

\author[0000-0003-3858-6361]{Shifan Zuo}
\affiliation{State Key Laboratory of Radio Astronomy and Technology, National Astronomical Observatories, CAS, A20 Datun Road, Chaoyang District, Beijing, 100101, P. R. China\\}

\correspondingauthor{Yougang Wang}
\author[0000-0003-0631-568X]{Yougang Wang}
\email{wangyg@bao.ac.cn}
\affiliation{State Key Laboratory of Radio Astronomy and Technology, National Astronomical Observatories, CAS, A20 Datun Road, Chaoyang District, Beijing, 100101, P. R. China\\}
\affiliation{School of Astronomy and Space Science, University of Chinese Academy of Sciences, Beijing 100049, China\\}
\affiliation{Key Laboratory of Cosmology and Astrophysics (Liaoning)\\
\& College of Sciences, Northeastern University, Shenyang 110819, China\\}

\author[0000-0001-6475-8863]{Xuelei Chen}
\affiliation{State Key Laboratory of Radio Astronomy and Technology, National Astronomical Observatories, CAS, A20 Datun Road, Chaoyang District, Beijing, 100101, P. R. China\\}
\affiliation{School of Astronomy and Space Science, University of Chinese Academy of Sciences, Beijing 100049, China\\}
\affiliation{Key Laboratory of Cosmology and Astrophysics (Liaoning)\\
\& College of Sciences, Northeastern University, Shenyang 110819, China\\}


\begin{abstract}

We present the \hi galaxy observation results of the FATHOMER (FAst neuTral HydrOgen intensity Mapping ExpeRiment), a pilot drift scan survey by the Five-hundred-meter Aperture Spherical radio Telescope (FAST). 
The survey comprises 28 hours of observations over 7 nights in 2021, covering a $60\, \deg^2$ sky area in the frequency range 1.05–1.45 GHz. The \hi galaxies are identified using both a matched-filtering algorithm and the SoFiA source-finding pipeline, which yield consistent detections. 
We derive the velocity width ($W_{50}$), flux density, and \hi mass for detected galaxies. 
A total of 702 galaxies are identified with \hi mass above $10^{6.2}\,{\rm M_\odot}$, signal-to-noise ratio greater than 5, and redshift $z < 0.09$. Among these, 331 are previously known from the ALFALFA survey. Of the newly detected sources, 9 have spectroscopic confirmation from SDSS, 285 are matched to SDSS or DESI photometric data, and 77 lack optical counterparts—possible candidates for dark or faint galaxies.
Comparison with ALFALFA shows that FAST enables detection of galaxies at higher redshifts and with lower \hi fluxes, despite the radio frequency interference (RFI) and partial data masking. 
A preliminary \hi mass function analysis reveals a higher characteristic mass and steeper low-mass slope than ALFALFA, indicating FAST’s enhanced sensitivity to massive and distant \hi systems. 
These results demonstrate FAST’s strong potential for future deep \hi surveys and highlight the importance of improved RFI mitigation and completeness correction.

\end{abstract}



\section{Introduction} \label{sec:intro}
Hydrogen is the most abundant element in the Universe, which is important in astronomical observations. The neutral hydrogen (\hi) 21cm spectral line is one of the key tools in radio astronomy research. 
After the epoch of reionization (EoR), 
\hi is widely distributed in galaxies, serving as an excellent tracer for studying the matter distribution, dynamics, and dark matter properties of galaxies. 
Analyzing \hi in galaxies can help us to understand the formation and evolution history of galaxies. 

A comprehensive wide-field \hi galaxy survey could provide valuable information for both cosmology and astrophysics research. Such surveys include 
the \hi Parkes All-Sky Survey \citep[HIPASS;][]{2001MNRAS.322..486B,2004MNRAS.350.1195M,2004MNRAS.350.1210Z}, the \hi Jodrell All-Sky Survey \cite[HIJASS;][]{2003MNRAS.342..738L}, and   
the Arecibo Legacy Fast ALFA (ALFALFA) survey \citep{2005AJ....130.2598G}. 
Other major \hi surveys such as 
the \hi-MaNGA programme, which provides \hi follow-up for the Mapping Nearby Galaxies at Apache Point Observatory survey \citep{2019MNRAS.488.3396M}, the Parkes Galactic All-Sky Survey (GASS; \citealt{2015A&A...578A..78K}) conducted with the Parkes 64-m Radio Telescope, the GALEX Arecibo SDSS Survey and it's extended GALEX Arecibo SDSS Survey (GASS/XGASS; \citealt{2010MNRAS.403..683C}), the \hi Nearby Galaxy Survey(THINGS; \citealt{2008AJ....136.2563W}) using the NRAO Very Large Array, the MeerKAT HI Observations of Nearby Galactic Objects: Observing Southern Emitters survey(MHONGOOSE;\citealt{2024A&A...688A.109D})and
the Widefield ASKAP L-band Legacy All-sky Blind surveY (WALLABY; \citealt{2020Ap&SS.365..118K}). 
In addition, several deep-field surveys have been conducted, such as the MeerKAT International GigaHertz Tiered Extragalactic Exploration survey (MIGHTEE;\citealt{2021A&A...646A..35M}), the Looking At the Distant Universe with the MeerKAT Array (LADUMA) survey \citep{2024AAS...24334607B}, the COSMOS HI Large Extragalactic Survey(CHILES;\citealt{2015AAS...22542703F}), the Arecibo Ultra-Deep Survey(AUDS;\citealt{2021MNRAS.501.4550X}) and the FAST Ultra-Deep Survey(FUDS;\citealt{2022PASA...39...19X}).


The recently built Five-hundred-meter Aperture Spherical Telescope (FAST), which is the largest single-dish radio telescope in the world \citep{2011IJMPD..20..989N},  has a great ability for \hi surveys. With its very high
sensitivity, more low-mass \hi objects can be observed in the local group~\citep{2018RAA....18....3L}, and more \hi galaxies at larger distances can be detected~\citep{2008MNRAS.383..150D,2019SCPMA..6259506Z,2020MNRAS.493.5854H}. Another advantage of FAST is that the L-band feed of FAST has 19 beams, which will greatly increase the speed of surveys. 

There are two ongoing large \hi surveys with the FAST: the Commensal Radio Astronomy FasT Survey (CRAFTS;~\citealt{2018IMMag..19..112L}), which  
is a multi-purpose large area drift-scan survey to 
observe Galactic and extragalactic \hi, as well as to search for pulsars and transients; and the FAST All Sky \hi Survey (FASHI; \citealt{2023arXiv231206097Z}), which is designed to cover the whole sky observable by FAST.  
Based on the data collected so far by the FASHI survey, some \hi galaxy catalogs have been released ~\citep{2022RAA....22f5019K,2023arXiv231206097Z}. 

In this paper, we focus on finding \hi galaxies in the FAst neuTral HydrOgen intensity Mapping ExpeRiment  \citep[FATHOMER;][hereafter L23]{Li2023}, which is a pilot drift scan survey for \hi intensity mapping. Compared with the studies by \citet{2022RAA....22f5019K} and \citet{2023arXiv231206097Z}, our work differs in several aspects. The survey region we selected is continuous and complete, located at the zenith of FAST site, where the antenna achieves its highest aperture efficiency \citep{2020RAA....20...64J}. This region was previously observed by the Arecibo telescope, providing a valuable reference for validating our results. 
We have independently developed our own data analysis pipeline {\tt fpipe}
\footnote{\url{https://github.com/TianlaiProject/fpipe.git}}. In \citetalias{Li2023}, we have shown the accuracy of our calibration strategy, which yields flux measurements of the radio continuum sources in excellent agreement with those found in the literature.

In the present paper, we will search for \hi galaxies in a completely surveyed area, and measure their \hi flux. We will also evaluate the detection capabilities for dwarf galaxies, and candidates of dark galaxies.
The study of dwarf and dark galaxies via \hi surveys plays a vital role in revealing the diversity of galaxy systems, especially at the low-mass end, and in refining theoretical models of galaxy formation and evolution\citep{2018RAA....18....3L}.

This paper is organized as follows: after this Introduction, in \refsc{sec:data}, we describe the FATHOMER observation and data products used in this study. \refsc{sec:analysies} introduces our methods for the blind searching of \hi galaxies. We present our main results and the discussions in \refsc{sec:results}.
Finally, we summarize this work in \refsc{sec:summary}. In this paper, we adopt a flat $\Lambda$CDM cosmology with parameters from the Planck 2018 results: $\Omega_{\mathrm{m}} = 0.315$, $\Omega_{\Lambda} = 0.685$, and $
\rm H_{0} = 67.4~\mathrm{km~s^{-1}~Mpc^{-1}}$ \citep{2020A&A...641A...6P}.

\section{Data} 
\label{sec:data}

\subsection{Observations}

\begin{table}
\caption{Summary of the technical details for the FATHOMER}
\label{table:par}
\hspace{-0.3cm}
{\small
\begin{threeparttable}
\begin{tabular}{ll}\hline\hline
Parameter       &   Value  \\\hline
R.A. range      &   $9^{\rm h}\leq \rm R.A. \leq 13^{\rm h}$\\
decl. range     &   $25.7^{\circ} \leq \rm decl. \leq 26.7^{\circ}$ \\
Receiver        &   L-band 19-beam array\\
Polarization    &   Linear dual-polarization (XX/YY)\\
Beam size       &   $2.95'$ at $1.4\,{\rm GHz}$\\
Gain\tnote{$\dagger$}   &   25.6 K Jy$^{-1}$\\
$T_{\rm sys}$               &   20 K\\
Frequency range         &   1300-1420 MHz\\
Spectral resloution&6.3 \kms\\
Spectral median rms&1.35 mJy at 6.3\kms resolution\\
\hline\hline
\end{tabular}
\begin{tablenotes}
\scriptsize
\item[$\dagger$] The gain of FAST could be expressed with $\eta G_{0}$, the $G_{0}$ value is 25.6K Jy$^{-1}$. The averaged gain values of 19 beams are listed in Table 5 of \citet{2020RAA....20...64J}.
\end{tablenotes}
\end{threeparttable}
}
\end{table}

\begin{figure*}
\centering
\includegraphics[width=\textwidth]{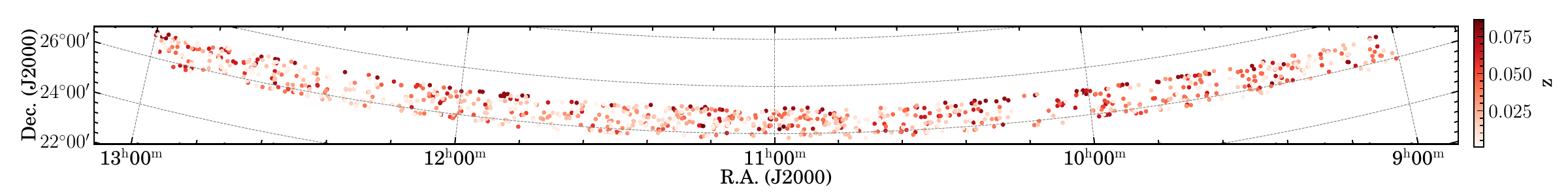}
\caption{The \hi galaxies detected within the FATHOMER field.}
\label{fig:sky}
\end{figure*}

We use the drift scan data that was carried out in March 2021 (as listed in \reftb{table:obs}).
The observed sky region, as shown in \reffg{fig:sky} with the detected \hi galaxies in this work, 
covers $\rm{09^h00^m<R.A.<13^h00^m}$, $\rm {25.7^{\circ}<decl.<26.7^{\circ}}$. 
 
During the observation, the telescope was fixed at a specific altitude along the celestial meridian, 
while the FAST feed array was rotated by $23.4^{\circ}$ relative to the compass points 
to enhance the declination coverage for each drift scan.
The noise diode is injected as a real-time calibrator for 0.9 s every 8 s. 
The time resolution of the calibrated time-ordered data (TOD)  is $1\,{\rm s}$. 
We use the Spec (W) spectrometer backend, which has 65,536 channels within the 500 MHz bandwidth, corresponding to the frequency resolution of $7.6$ kHz. 

\begin{table}
\begin{center}
\caption{Observation mode. 
Column (1): The observational sky area's field center.
Column (2): The observation date. }
\label{table:obs}
\begin{tabular}{cc}\hline\hline
Field center    &     Date   \\\hline    
HIMGS 1100+2600 & 2021-03-02          \\
HIMGS 1100+2632 & 2021-03-05         \\
HIMGS 1100+2643 & 2021-03-06         \\
HIMGS 1100+2654 & 2021-03-07        \\
HIMGS 1100+2610 & 2021-03-09        \\
HIMGS 1100+2621 & 2021-03-13         \\
HIMGS 1100+2610 & 2021-03-14          \\
\hline\hline
\end{tabular}
\end{center}
\end{table}

We took the 3C286 as our flux calibrator and performed the calibration observation after
each 4-hour drift scan survey. 
The calibration was also performed in drift-scan mode, with the 19 feeds arranged into 
five east–west lines. By employing five different drift scans, the calibrator drifted 
across each feed.

\subsection{Map-making}\label{sec:todpipeline}

From the time-ordered data, which include the readout of different feed beams, we can make a 3D image cube of the surveyed region, with intensity as a function of both direction and frequency. Subsequently, the \hi galaxy searching can be conducted in the map domain, i.e. from this data cube. We adopted the {\tt fpipe}, which was presented in our previous work \citepalias{Li2023}. Here
we give a brief summary of the data processing pipeline and refer the interested reader to \citetalias{Li2023} for the details: 

\begin{enumerate}
\item The raw FITS files are converted to HDF5 files, which combine the time-ordered data of 19 beams as an extra axis and split the full frequency band into three sub-bands, i.e., the low-frequency band $1050$--$1150$ MHz, the mid-frequency band $1150$--$1250$ MHz, and the high-frequency band $1250$--$1450$ MHz.
The pointing directions of every time stamp are calculated and recorded together with the data in the HDF5 file.
\item The time stamps with the noise diode on are extracted and used to estimate the bandpass gain factor $g(\nu)$, and the temporal gain drift factor $g(t)$. An independent and strict RFI flagging process is applied to the noise-diode-on data.
\item The estimated bandpass gain factor $g(\nu)$ is then applied to the noise-diode-free data to correct the frequency response.
It is worth noting that $g(\nu)$ is normalized, and the bandpass-corrected data remain dimensionless. This correction aims to eliminate strong 
frequency-dependent variations, thereby facilitating more effective RFI flagging in the following step.
\item The {\tt SumThreshold} and {\tt SIR} (Scale-Invariant Rank) RFI flagging process \citep{2021A&C....3400439Z} are then applied to the bandpass calibrated data.
\item Apply the temporal and the absolute gain factors to the RFI-flagged data.
The absolute gain calibration is done by rescaling the noise-diode temperature with the celestial calibrator, which is observed once per night.
\item In addition to the gain variation, there is still a noise level offset, which 
varies over time and between different feeds and days.
The data is then zero-centered by subtracting the temporal noise level variation.
\item The flux-calibrated and zero-centered data is finally passed to the map-making process.
We adopt a kernel-weighted map-making procedure developed in \citetalias{Li2023} with 
a Gaussian kernel,
\begin{align}
K_{pq} = \exp\left[-\frac{1}{2}\left(\frac{r_{pq}}{\sigma_{K}}\right)^2\right],
\end{align}
where $r_{pq}$ is the great-circle distance between the $p$th and $q$th pixels of the map,
and $\sigma_K = 1.5'$ indicates the kernel size. 
The map is constructed using the HEALPix scheme with \texttt{NSIDE}=2048, 
corresponding to a pixel size of $1.72'$. The data from different days and feeds are
directly averaged on the map domain.

\end{enumerate}

\section{\hi Galaxy Searching Methods}\label{sec:analysies}

In this paper, we use two blind searching methods to search for the \hi galaxies,
i.e., the matched-filter algorithm  developed in the ALFALFA survey\citep{2007AJ....133.2087S}, and SoFiA \citep{2021MNRAS.506.3962W}. 
Both methods can be applied to the map. In addition, we also applied the 
matched-filter method to the initial time-ordered data to avoid false identification.
The detailed discussions are presented in \refsc{subsec:selection criteria}.

\subsection{Matched-Filtering Method} \label{subsec:macthed-filter}

Compared with the conventional peak-finding adopted in the HIPASS~\citep{2001PhDT.......230K},
the matched-filter method is more sensitive to the total flux of the \hi galaxy, 
which could enhance the sensitivity and efficiency, potentially allowing for the detection of 
fainter \hi signals.  
The matched-filter method has been applied to the FAST observation data, e.g.,
CRAFTS and FASHI, for \hi absorption searching \citep{2023A&A...675A..40H,2025ApJS..277...25H}.
In this method, the spectrum is cross-correlated with a set of \hi profile templates,
\begin{align}
c(\delta\nu) =& S(\nu) \star T(\nu - \delta\nu; \theta) \nonumber \\
       =& \frac{1}{n_\nu \sigma_S \sigma_T} \sum_{\nu} \left(S(\nu)T(\nu-\delta\nu;\theta)\right),
\end{align}
where $c(\delta\nu)$ represent the cross-correlation function with frequency interval of $\delta\nu$,
$S(\nu)$ is the spectrum measurement extracted from the map,
$T(\nu; \theta)$ represents the \hi profile template with $\theta$ denoting its parameter sets.
In addition, $n_\nu$ is the number of frequency channels considered in matched filter processing, 
$\sigma^2_S \equiv \frac{1}{n_\nu} \sum_{\nu} S^2(\nu)$ is the variance of the measured spectrum,
and $\sigma^2_T \equiv \frac{1}{n_\nu} \sum_{\nu} T^2(\nu)$ is the variance of the template, respectively.
The $\chi^2$ function is defined as,
\begin{align}\label{eq: chi2}
\chi^2 = n_\nu \sigma^2_S \left( 1 - c^2(\delta\nu)\right).
\end{align}
The minimum $\chi^2$ is achieved by maximizing the cross-correlation function. 
The \hi galaxy candidates are searched by varying the template parameters to achieve the
maximized cross-correlation function.
We refer the interested reader to \citet{2007AJ....133.2087S} for further details.
In this work, we adopt a simple Gaussian function as the template function,
\begin{align}
T(\nu; T_{0}, \nu_0, \sigma_\nu) = T_{0} \exp\left(-\frac{1}{2}\frac{(\nu-\nu_0)^2}{\sigma^2_\nu}\right),
\end{align}
where the free parameters $T_0$ as the overall amplitude, $\nu_0$ as the central frequency,
and $\sigma_\nu$ as Gaussian kernel size. 
We adopt a single-Gaussian template for \hi line detection due to its symmetry, minimal parameterization, and computational efficiency \citep{2014PASA...31...23W}. This template enables fast computation during the template-matching or filtering steps. Although some spectral profiles exhibit non-Gaussian features (e.g. asymmetry, multiple peaks), a single Gaussian template nevertheless effectively identifies the line center and integrated flux. Additionally, we employ the first two symmetric Hermite functions to fit \hi profiles of varying widths. 
In order to identify the \hi galaxies with different sizes, the \hi profile
width $\sigma_\nu$ is varying within a large enough parameter space
corresponding to the velocity width range from $6.3\,{\rm km\,s}^{-1}$ to $253.5\,{\rm km\,s}^{-1}$
with step-size of $6.3\,{\rm km\,s}^{-1}$.

The \hi galaxies are then identified according to the ${\rm S/N}$ of the profile
smoothed with a kernel size of $W_{\rm K} = 400\,{\rm km\,s}^{-1}$(where typical baseline fluctuations start to be of the same width as the galaxy profiles), 
which is defined as,
\begin{eqnarray}
  {\rm S/N} =
    \begin{cases}
      \frac{S_{21}/W_{50}}{\sigma_{\mathrm{rms}}} \sqrt{\frac{W_{50}/2}{\Delta v}}&
      W_{50} < W_{\rm K} \\
      \frac{S_{21}/W_{50}}{\sigma_{\mathrm{rms}}} \sqrt{\frac{W_{\rm K}/2}{\Delta v}}& 
      W_{50} \geqslant W_{\rm K}\\
    \end{cases},
    \label{signal_to_noise}
\end{eqnarray}
where $W_{50}$ is the profile velocity width of the signal, 
$S_{21}$ is the integrated flux across the full profile width,
$\sigma_{\mathrm{rms}}$ is the rms noise figure across the spectrum measured in $\mJy$ 
at velocity resolution of $\Delta v = 6.3$ \kms.

\begin{figure}
\centering
\includegraphics[width=0.49\textwidth]{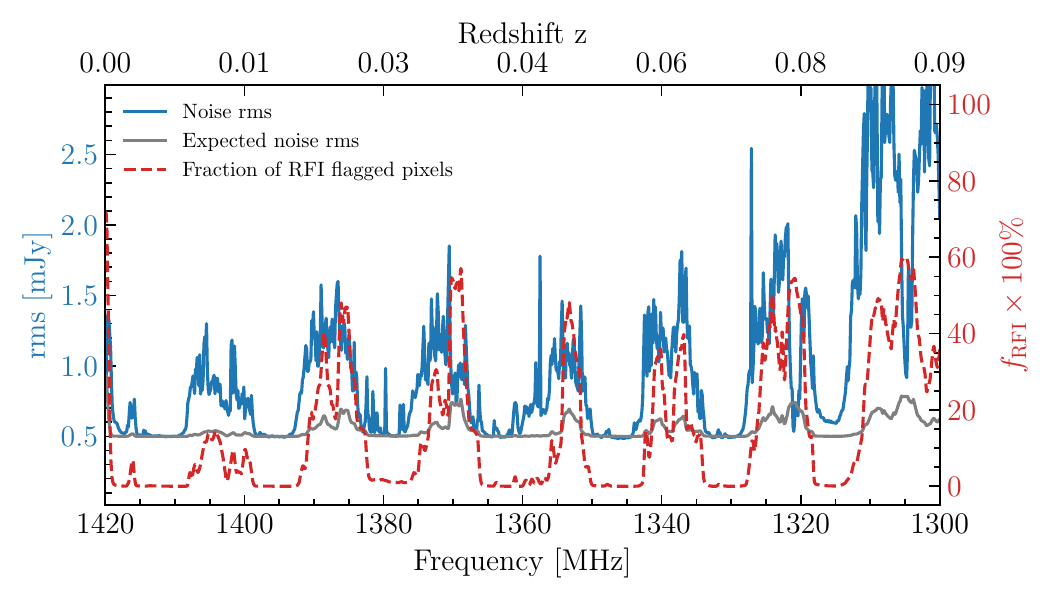}
\caption{
Frequency dependence of the noise level (blue curve, left axis)
and the corresponding fraction of RFI-masked pixels (red dashed curve, right axis). 
The bottom axis shows the observed frequency in MHz, while the top axis indicates the corresponding redshift of 
the \hi line. 
The secondary y-axis illustrates the fraction of RFI-masked pixels over the same frequency range.
The gray curve represents the expected rms increase induced by RFI flagging, which is estimated 
via \refeq{eq:rmsexp}.
}
\label{fig:rms}
\end{figure}

The noise level, $\sigma_{\mathrm{rms}}$, varies across the frequency band
due to frequency-dependent data masking and residual RFI. 
To estimate the systematic rms, we compute the frequency differential
of four adjacent channels in the observed map \citepalias[for details, see][]{Li2023},
thereby suppressing contributions from continuum sources. 
The noise level in each frequency channel is then derived from the 
standard deviation of the continuum-subtracted map. 
The resulting rms spectrum is shown in \reffg{fig:rms} with the blue curve (left axis).
In addition, the corresponding fraction of RFI-flagged pixels, $f_{\rm RFI}$, as a function 
of frequency is shown with the red dashed curve (right axis). 
Enhanced rms values are evident at certain frequencies, 
corresponding to channels that are heavily contaminated by RFI.
The gray curve represents the expected rms increase induced by RFI flagging, as given by,
\begin{align}\label{eq:rmsexp}
\sigma_{\rm rms, exp} = \sigma_{\rm rms, RFI\text{-}free} / \sqrt{1-f_{\rm RFI}},
\end{align}
where $\sigma_{\rm rms,,RFI\text{-}free} = 0.5,\mathrm{mJy}$ is the mean rms level measured in 
RFI-free frequency ranges. 
We find that the expected rms systematically underestimates the measured values. 
This discrepancy indicates the presence of significant residual RFI, 
which introduces additional noise and leads to an excess rms beyond the prediction based solely on data loss from flagging.

The matched-filter method is applied to both the initial TOD and the map pixel-by-pixel.
When we search for \hi candidates, the matched-filter method is performed for each pixel. Once a candidate is detected, the parameters such as signal-to-noise ratio and  \hi flux are calculated for the surrounding eight pixels. Candidates with adjacent pixels and the central frequency difference smaller than 0.2 MHz will be merged to one candidate.  

To enhance the signal-to-noise ratio of the \hi candidates, we record only the total flux in the final maps, rather than the fluxes from the two orthogonal polarizations ($S_{XX}$ and $S_{YY}$). The validity of the \hi candidates obtained in the mapping data is then checked using the TOD. A real \hi galaxy exhibits nearly zero polarization. Furthermore, it should not be detectable simultaneously by multiple beams (unless at very low redshift), nor should its signal persist for an anomalously long time within a single beam.   

For the searching with TOD, the matched-filter method is applied to the 
$S_{XX}$ and $S_{YY}$ polarizations, as well as the combined Stokes $I = S_{XX} + S_{YY}$ data,
individually. We set the threshold ${\rm S/N} > 3.5$ for the polarized data $S_{XX}$ and $S_{YY}$
and  ${\rm S/N} > 5.5$ for the combined $I$ data, respectively.      
The candidate selection criteria, based on both TOD and mapping data, are detailed in Section~\ref{subsec:selection criteria}.


In this work, we only use the Gaussian kernel for the matched-filter galaxy searching.
However, the Gaussian kernel fits well with the galaxy \hi profile of single-horn shape,
and may result in bias in \hi profile parameter extraction for double-horn shape.
Therefore, the matched-filter method is only used for galaxy candidate searching. 
The profile parameters inference is carried out by fitting with a busy function 
\citep{2014MNRAS.438.1176W}. Details of the profile parameters inference are further 
discussed in \refsc{sec:results}.

\subsection{\sofia} \label{subsec:sofia}
\sofia is a flexible pipeline designed to search and parametrize the \hi galaxy 
in a three-dimensional data cube \citep{2015MNRAS.448.1922S}. 
Recently, the \sofia has been improved to \sofiaii \citep{2021MNRAS.506.3962W}. Compared to its predecessor, \sofiaii is significantly faster and more memory-efficient.

In this work, we adopted one of the \sofiaii algorithms, {\tt Smooth+clip} algorithm \citep{2012MNRAS.422.1835S}. The {\tt Smooth+clip} galaxy searching 
is carried out by finding pixels over a pre-chosen threshold, with the map smoothed 
by multiple user-defined kernel sizes.

We choose the {\tt Smooth+clip} threshold to be $3.5$ times the local noise rms of the data cube. 
We adopted a set of 3-dimensional Gaussian kernels
,
\begin{align}
K(\Theta, v) = \exp\left(-\frac{1}{2}\left(\frac{(\Theta/\Delta\Theta)^2}{\sigma^2_\perp} + 
\frac{(v/\Delta v)^2}{\sigma^2_\parallel} \right) \right),
\end{align}
where $\Theta$ and $v$ represent the angular separation and the velocity differences 
with respect to the kernel center, and $\Delta \Theta$ and $\Delta v$ are the 
pixel size and the velocity resolution of the map,
respectively. $\sigma_\perp$ and 
$\sigma_\parallel$ are the user-defined kernel sizes in the direction perpendicular and parallel
to the line-of-sight. 
In this work, we set $\sigma_\perp \in \left\{0, 3,5, 10\right\}$ 
and $\sigma_\parallel \in \left\{0, 3, 7, 15, 31, 40\right\}$.
The kernel parameter setting may need slight adjustments based on the
actual number of pixels across the beam and the spectral resolution of the cube.
As shown in ~\cite{2021ascl.soft09005W}, SoFiA2 will reject the linked sources that fail to satisfy the size criteria set by the user, which can help to remove potentially false detections from the noise spikes or artifacts. The linker (module linker) allows us to merge detected pixels into coherent sources.
We choose to link pixels across a spatial radius of 2 and a spectral radius of 2 here.
The linking parameters specify that the minimum size of sources in the spatial dimension is 5 pixels per channel in both XY and Z space, while the maximum size in XY space and Z space is default.
In this case, the effective upper limit depends on the dimensions of the input data cube. The maximum linking length was set by the dimensions of the cube.

\subsection{Candidate Selection}\label{subsec:selection criteria}
The RFI is a major challenge for searching the \hi galaxy~\citep[e.g.][]{2010MNRAS.405..155O,2023RAA....23k5006L}. 
As described in \refsc{sec:todpipeline}, we have applied the {\tt SumThreshold} and {\tt SIR} RFI flagging 
process to the TOD.
However, due to the complexity of RFIs, the flagging results strongly depend on the threshold. 
It may result in RFI residuals with a permissive threshold and, conversely result in 
galaxy \hi signal over-subtraction with a stringent threshold.
Residual RFI may occasionally be misidentified as an \hi galaxy by our source-finding algorithm. 
However, leveraging the multi-beam and dual-polarization measurements from the TOD allows us 
to further discriminate and eliminate such false detections. 

With the galaxy candidates found in the map, we extract the corresponding TOD according to  
their coordinates.
A candidate will be removed as an RFI, if its corresponding TOD meets any of the following conditions. 
\begin{enumerate}
\item It is detected in multiple beams simultaneously. If the candidate is detected in more than 9 beams simultaneously, it is directly removed. The reason is that we find the galaxy with the largest angular size in our sample is nearby galaxy AGCNr7772 (NGC 4565), which has been detected in nine beams centered on Beam 8, with flux density and S/N declining by approximately 50\% in the outermost beams. The parameter of $\rm W_{50}$ for this galaxy is  509 km/s. Similar beam-dependent attenuation patterns were found in other bright, nearby galaxies with broad profiles.
Based on these results, we reject all candidates appearing in more than nine beams. 
If the candidate is detected in 2 - 9 beams simultaneously, we further inspect the 
detecting beams and their redshift. We examined the spatial distribution of beams where the candidate signal was detected. Specifically, we calculated the maximum projected angular separation ($\theta_{\mathrm{project}}$) between the beams.
When this separation exceeds 6 arc minutes and exhibits a relatively high redshift($z\ge0.06$), the signal is considered likely spurious and more likely to be caused by interference or other artifacts. For those detected in 2–9 beams with redshift z$<$ 0.06, we retain the source if the flux peak and S/N in the outermost beam decrease by more than 50\% relative to the beam with the largest flux density. Common RFIs typically appear in a majority of beams with little or no attenuation.
This selection strategy effectively suppresses RFI while preserving genuine extragalactic HI sources, particularly those with narrow profiles and higher redshifts, striking a balance between sensitivity and reliability.

\item 
Candidates appearing in more than 60 continuous or discontinuous time samples(time resolution is 1s) in the TOD data of a single beam are further examined. Some nearby galaxies with broad profiles can span over 60 time points during source extraction. If the S/N peaks and then gradually declines within this window, the signal is likely real. In contrast, if the S/N remains above 5 without a clear decay, the candidate is more likely RFI.

\item A significant difference between the polarized measurements $S_{XX}$ and $S_{YY}$,
i.e., their \hi profile peak frequencies differ more than $0.5$ MHz or their profile amplitudes differ
more than $30\%$.
\end{enumerate}

During the candidate selection process for our galaxy survey, we initially identified several extremely massive galaxy. Among these, two candidates were traced to beam 15. Further 
investigation revealed that this beam suffered from instrumental faults, which 
produced a spurious signal in one day of observations that closely mimicked the 
spectral profile of a genuine galaxy. For the remaining extremely massive candidates, although they were not evident in the spectra of a snapshot, they
were clearly identified as RFI in the corresponding waterfall plots.
To ensure the robustness of the final sample, each \hi candidate was independently re-evaluated by two team members through a comprehensive visual inspection, including the imaging products, 
spectral profiles (from both the map-domain data and TOD), and waterfall plots.
Following this rigorous cross-validation, all spurious signals were removed from the sample.

\section{Results} \label{sec:results}

\subsection{Extragalactic \hi catalog}
 
\begin{table*}
\begin{center}
\caption{Examples of the extragalactic galaxy catalog.}
\label{tab:gal}
\hspace*{-1.5cm}
{\scriptsize
\begin{tabular}{ccccccccccc}\hline\hline
Galaxy ID & \hi centroid & OC centroid & $cz_{\odot}$ & $W_{50}$ & S/N & $S_{21}$  & $\sigma_{\rm rms}$ & $D_{\rm L}$ & $\log M_\hi / M_\odot$ & Flag \\
          & J2000  & J2000  & \kms &\kms & &
\Jy$\cdot$\kms & \mJy & \Mpc &  &   \\\hline
1 & 090032.40+255736.00 & 090032.80+255651.00 & 14452 &   70 &  9.6 &  0.79 & 2.25 &     233.0 &    9.98 &     {\tt Code 1} \\
  2 & 090232.93+262406.05 & 090226.75+262346.18 &  3808 &  117 &  7.9 &  0.50 & 1.78 &      57.6 &    8.75 &    {\tt Code 3} \\
  3 & 090238.70+255611.00 & 090238.60+255602.00 &  2428 &  356 & 90.3 & 14.96 & 1.86 &      36.4 &    9.66 &    {\tt Code 1} \\
  4 & 090309.84+261251.95 & 090309.67+261110.72 & 11576 &  249 &  8.5 &  0.20 & 1.49 &     183.4 &    9.20 &    {\tt Code 3} \\
  5 & 090325.66+261406.79 & 090332.19+261212.60 & 12407 &  139 &  5.0 &  0.20 & 1.41 &     197.5 &    9.26 &    {\tt Code 3} \\
  6 & 090418.40+262635.99 & 090422.83+262607.62 & 13032 &   98 & 11.1 &  0.69 & 1.87 &     208.3 &    9.83 &    {\tt Code 3} \\
  7 & 090455.31+260523.14 & 090456.59+260621.60 & 11750 &  118 &  6.4 &  0.34 & 1.22 &     186.3 &    9.43 &    {\tt Code 2} \\
  8 & 090455.31+261751.40 & 090458.93+261745.60 & 12394 &   96 &  5.8 &  0.24 & 1.43 &     197.3 &    9.33 &    {\tt Code 3} \\
  9 & 090458.30+260651.01 & 090456.60+260622.00 & 11710 &   70 &  6.9 &  0.34 & 1.83 &     185.6 &    9.44 &     {\tt Code 1} \\
 10 & 090505.86+263521.26 &                     & 11293 &  163 & 10.2 &  0.66 & 1.63 &     178.6 &    9.88 &     {\tt Code 4} \\
 11 & 090613.50+254639.00 & 090612.40+254701.00 &  2595 &  138 &  9.3 &  1.48 & 2.59 &      38.9 &    8.35 &   {\tt Code 1} \\
 12 & 090630.24+260523.14 & 090627.72+260643.67 &  5322 &  120 &  5.1 &  0.27 & 1.12 &      81.2 &    8.62 &    {\tt Code 3} \\
 13 & 090738.79+255410.80 & 090739.11+255646.00 &  5315 &  118 &  6.2 &  0.50 & 1.64 &      81.1 &    8.88 &    {\tt Code 3} \\
 14 & 090831.52+260408.36 & 090838.74+260317.21 & 18835 &   87 &  8.8 &  0.21 & 0.91 &     311.9 &    9.83 &    {\tt Code 3} \\
 15 & 090912.40+263551.00 & 090908.30+263611.02 &  2786 &  204 & 17.8 &  1.76 & 1.67 &      41.8 &    9.17 &   {\tt Code 1} \\
 16 & 090935.40+263628.01 & 090934.80+263633.01 &  2752 &   88 & 35.0 &  3.09 & 2.27 &      41.3 &    9.25 &    {\tt Code 1} \\
 17 & 090937.54+263521.16 & 090943.82+263631.93 &  9682 &  162 & 22.1 &  1.55 & 1.45 &     151.6 &    9.91 &    {\tt Code 3} \\
 18 & 090940.08+261646.60 & 090939.90+261753.95 & 22056 &   91 &  7.5 &  0.30 & 1.49 &     375.4 &    9.85 &    {\tt Code 2} \\
 19 & 090946.70+265129.99 & 090947.80+265138.02 &  6327 &  125 &  7.9 &  1.00 & 3.19 &      97.1 &    9.87 &    {\tt Code 1} \\
 20 & 090952.57+261751.40 & 090954.74+261952.97 & 22056 &   57 &  5.5 &  0.21 & 1.22 &     372.7 &    9.79 &    {\tt Code 3} \\
 21 & 090958.30+270031.00 & 090958.60+270011.02 &  6327 &  131 & 18.9 &  2.80 & 2.47 &      97.1 &    9.79 &    {\tt Code 1} \\
 22 & 091038.09+264136.85 &                     & 13051 &  121 &  9.0 &  0.98 & 1.52 &     208.6 &    9.98 &    {\tt Code 4} \\
 23 & 091040.30+262825.00 & 091038.10+262832.02 & 13329 &  307 &  5.1 &  0.93 & 2.33 &     203.0 &    9.96 &   {\tt Code 5} \\
 24 & 091146.64+265253.80 &                     &  9180 &  126 &  7.4 &  0.29 & 2.31 &     143.3 &    9.14 &    {\tt Code 4} \\
 25 & 091311.02+265253.80 &                     &  1419 &   96 &  7.1 &  0.29 & 1.37 &      21.1 &    7.83 &   {\tt Code  4} \\
 26 & 091316.29+265409.07 &                     & 15080 &   88 &  8.5 &  0.29 & 1.73 &     244.0 &   10.16 &    {\tt Code 4} \\
 27 & 091346.60+260430.00 & 091347.00+260450.99 & 14425 &  331 &  7.5 &  0.82 & 1.42 &     232.5 &    9.98 &    {\tt Code 1} \\
 28 & 091356.00+270011.99 & 091356.40+270027.00 &  7699 &  117 & 11.4 &  2.00 & 3.15 &     119.1 &    9.81 &   {\tt Code 1} \\
 29 & 091440.66+265138.52 & 091449.70+265211.17 & 15087 &   99 &  6.7 &  0.98 & 1.64 &     241.5 &   10.13 &   {\tt Code 3}\\
 30 & 091442.10+264153.02 & 091439.30+264156.00 & 11723 &  111 &  6.3 &  0.37 & 1.42 &     185.9 &    9.70 &    {\tt Code 1} \\
......\\
\hline\hline
\end{tabular}
}
\end{center}
\end{table*}

\begin{table}
\caption{Summary of \hi galaxy (candidate) detection with the FATHOMER.}\label{table:catalog}
{\centering
\hspace{-1.0cm}
\begin{tabular}{cccc}\hline
Cross-matched Survey & Matched-filtering & SoFiA & Flag \\ \hline
\multicolumn{4}{c}{Confirmed \hi galaxies} \\
ALFALFA              & 331               & 331   & {\tt Code 1}\\
Missed due to RFI    & 120               & 120   & {\tt Code 5}\\ \hline
\multicolumn{4}{c}{Newly detected \hi galaxy candinates} \\
SDSS spectroscopic   & 9                & 9    & {\tt Code 2}\\
SDSS and \textbf{DESI} photometry      & 285               & 259   & {\tt Code 3}\\
No matched OCs       & 77               & 70   & {\tt Code 4}\\
\hline\hline
\end{tabular}
}
\end{table}
Unless stated otherwise, the results presented in the paper are from the matched-filtering approach.
There are 702 \hi galaxy candidates identified within the observed sky area.
Table~\ref{tab:gal} shows examples of the \hi galaxy detected in our selected region\footnote{The full catalog is available at \url{https://doi.org/10.5281/zenodo.18782636}}. 
The content of each column is as follows:
\begin{itemize}[leftmargin=0.4cm]
\item Column 1: Index number of the galaxies identified in the FATHOMER survey area.

\item Column 2: The \hi galaxy centroid coordinate in J2000 Equatorial system (formatted as {\tt hhmmss.sSddmmss}), with the pointing accuracy of $\sim 7.9''$ \citep{2020RAA....20...64J}. 

\item Column 3: The associated optical counterpart (OC) centroid coordinate in J2000 Equatorial system (formatted as {\tt hhmmss.sSddmmss}). The cross-identification of OC is discussed later in the content
of Column 11 {\tt Code 2}.

\item Column 4: Heliocentric velocity of the \hi galaxy $cz_\odot$ in \kms, measured as the midpoint
of the \hi velocity profile. 
The central velocity of each source is given with the matched-filter method by minimizing 
the $\chi^{2}$ in the \refeq{eq: chi2}.

\item Column 5: The velocity width of the \hi line profile $\rm W_{50}$ in $\rm km\ s^{-1}$, measured at the 50$\%$ level of peak flux. The velocity width is extracted by fitting the \hi profile with a
busy function developed in \citet{2014MNRAS.438.1176W} via a public package 
BusyFit\footnote{\url{https://gitlab.com/SoFiA-Admin/BusyFit}}.

\item Column 6: The detection S/N, estimated by Eq.~(\ref{signal_to_noise}). 

\item Column 7: The integrated \hi flux $S_{21}$ in ${\rm Jy}\,{\rm km\,s}^{-1}$ 
measured by the matched-filter approach. 
The \hi galaxy is extracted by summing over the \hi flux in all the galaxy pixels at each frequency channel \citep{2018ApJ...861...49H},
\begin{equation}\label{HI spectrum summed}
S_{21}(\nu) = \frac{\sum_{\Theta} S_{\nu}(\Theta)}{\sum_{\Theta} B(\Theta)},
\end{equation}
where $S_{\nu}(\Theta)$ is the flux density in each pixel, and $B(\Theta)$ is the beam values
at the position $\Theta$, where $\Theta$ represents the angular
separation with respect to the beam center.
We set up a pixel where the source is located, along with the surrounding 8 pixels, which is large enough to cover the beam-convolved galaxy size. 
The total \hi flux is integrated over the frequency region where the \hi signal is present, with an integration width of $1.8 W_{50}$.

\item Column 8: The noise rms across the spectrum $\sigma_{\rm rms}$ measured in $\mJy$ at
velocity resolution of 6.3 \kms.

\item Column 9: Luminosity Distance $D_{\rm L}$ in Mpc corresponding to the frequency at the midpoint of the HI velocity profile.
We assume the $\Lambda$CDM cosmology model and adopt 
cosmology parameters of Planck 2018 \citep{2020A&A...641A...6P}.

\item Column 10: Logarithm of the \hi mass, $\log M_{\hi} / M_\odot$. 
We derive the \hi mass via \citep{2015A&ARv..24....1G}, 
\begin{eqnarray}
    \frac{M_{\hi} }{M_{\odot } } = \frac{2.35\times 10^{5}}{1+z}\left(\frac{D_{\rm L}}{\rm Mpc}\right)^2
    \left(\frac{S_{21}}{{\rm mJy\cdot km\,s}^{-1}}\right)
    \label{HImass function}
\end{eqnarray}
where the distance $D_{\rm L}$ is the luminosity distance given in Column 9 and 
$S_{21}$ is the total \hi flux given in Column 7. 

\item Column 11: The relevant coded flags indicate different \hi detections defined as follows. 

{\tt Code 1} refers to sources that have been observed by both ALFALFA ~\citep{2018ApJ...861...49H} and FATHOMER. These sources have $\rm S/N$$\geqslant 5$. As shown in ~\citep{2018ApJ...861...49H}, sources with $\rm S/N$$\geqslant6.5$ are nearly 100$\%$ reliable. 
Although some candidates  with Code 1 in Table 2 have $\rm S/N$ $<$6.5, they have already been matched with their OCs in \cite{2018ApJ...861...49H}. 

{\tt Code 2} refers to some new sources found by FATHOMER, which have been matched with OCs with known optical redshifts obtained from the spectrum of the galaxies. We do cross-identification of the FAST detection sources with the optical galaxies in the Sloan Digital Sky Survey (SDSS;\citealt{2009ApJS..182..543A}) DR7 database. The matched OC should have the shortest angular distance within three arcminutes around the \hi galaxy in FATHOMER and the central frequency difference $|\nu_f-\nu_o|<0.2\,{\rm MHz}$, here $\nu_f$ and $\nu_o$ represent the central frequencies of the galaxy obtained in the FATHOMER and in the SDSS redshift, respectively.

{\tt Code 3} refers to some FAST newly found sources, which have been matched with OCs with uncertain optical redshifts, i.e., these OCs have no spectroscopic redshifts but have photometric redshifts. 
We do cross-identification of these sources with the SDSS Photoprimary database 12 ~\citep{Alam_2015} and DESI Image Legacy Surveys DR9 \footnote{\url{https://gax.sjtu.edu.cn/data/DESI.html}}. 
For each \hi candidate, we calculate the angular distances $d_A$
to all photometric objects within 3 arcminutes of it. For each photometric galaxy, the photometric redshift is $z \pm \sigma_{z}$, where $\sigma_{z}$ is the redshift error. From the central frequency of the \hi candidate, we can get the redshift $z_f$ from the FAST observation. A possible OC of \hi galaxy should satisfy  $(z-\sigma_z)\leq z_f\leq (z+\sigma_z)$. We then calculate the ratio between the distance $d_A$ and the Petrosian radius $r_p$ of the optical galaxy within three arcminutes. We take the galaxy with $(z-\sigma_z)\leq z_f\leq (z+\sigma_z)$ and the minimum value of $d_A/r_p$ as the OCs.       

{\tt Code 4} refers to some FAST newly found sources, which have no OCs by cross-matched with the SDSS and DESI photometric catalogs. If there is no OC within 3 arcminutes around the \hi candidate that satisfies $(z-\sigma_z)\leq z_f\leq (z+\sigma_z)$, then we called these \hi candidates as no OCs. These candidates are likely to be dwarf or faint galaxies.    

{\tt Code 5} refers to sources that have been  observed by ALFALFA and been contaminated by the strong RFIs in FATHOMER.

\end{itemize}

In Table~\ref{table:catalog}, we show the summary of the \hi galaxy detection for
using both the matched-filter and the SoFiA method. 

\subsection{Flux measurements comparing}

\begin{figure*}
\centering
\includegraphics[width=\textwidth]{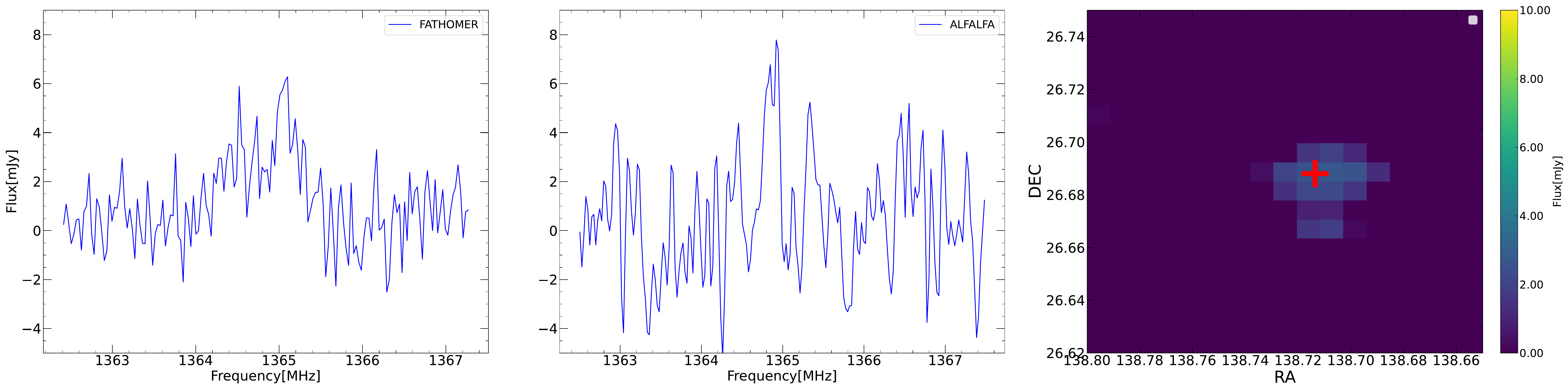}
\caption{Frequency spectra for galaxy AGCNr 191363 in FATHOMER (left) and ALFALFA (middle). The right panel is the sky map in FATHOMER. The signal in ALFALFA is relatively weak, with $\rm S/N=4.3$. In FATHOMER, we get a higher S/N ($\rm S/N=6.3$) and confirm that this signal should be real. 
}
\label{fig:191363}
\end{figure*}

\begin{figure}
\centering
\includegraphics[width=0.47\textwidth]{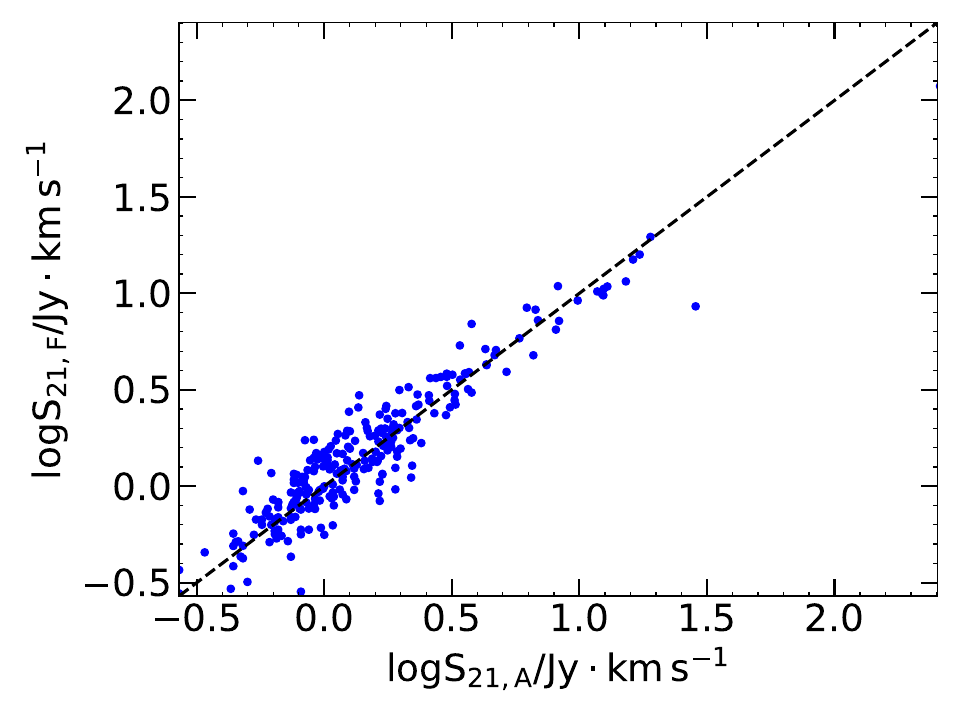}
\includegraphics[width=0.47\textwidth]{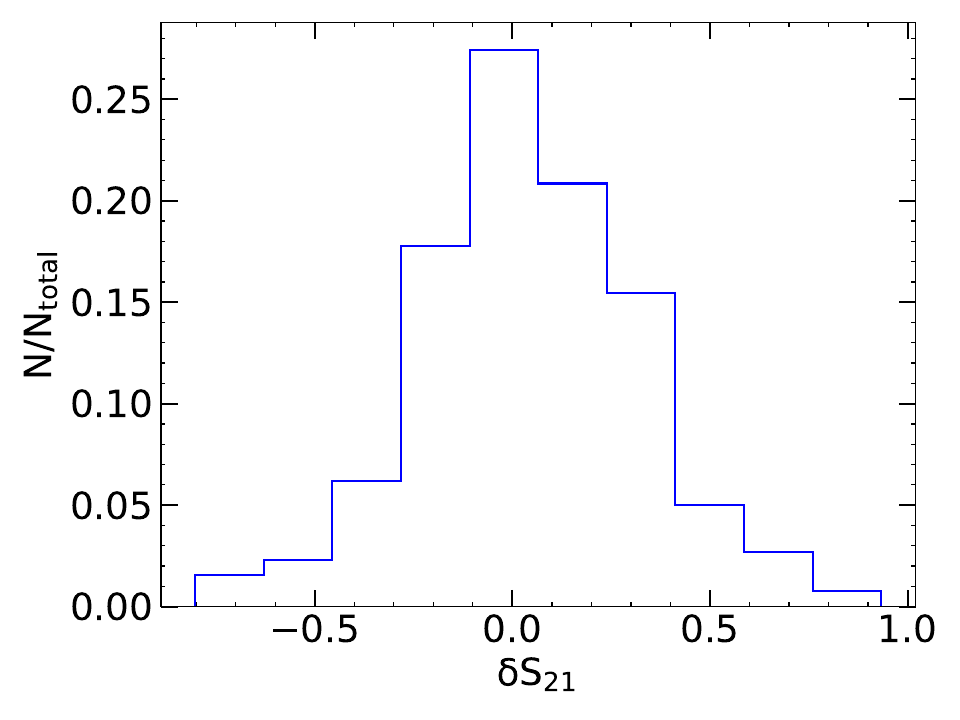}
\caption{{\it Top}: Comparison of the flux between FATHOMER and ALFALFA. The black dashed line represents the equality of two fluxes from two different surveys. All galaxies shown here have $\rm S/N>10$.
{\it Bottom}: Histogram of $\rm \delta S_{21}$.
}
\label{fig:hiflux}
\end{figure}

Since the FATHOMER survey area overlaps with that of the ALFALFA survey, a comparison between the galaxies
detected by the two surveys can help verify the reliability of FATHOMER detections.

We show an example of the measured spectrum and image of \hi galaxy AGCNr 191363
in \reffg{fig:191363}, where the spectra measured with FATHOMER and ALFALFA are
shown in the left and middle panels, and the image measured with FATHOMER is
shown in the right panel.
This galaxy is also found in ALFALFA with $ \rm S/N_{A} = 4.3$, 
where ${\rm S/N_{A}}$ denotes the S/N defined with ALFALFA nosie level. 
From the spectrum of FATHOMER map, we get $\rm S/N_{F} = 6.3$,
where $\rm S/N_{F}$ denotes the S/N defined with the FAST noise level.

To compare the flux measured by FATHOMER and ALFALFA, we use the \hi galaxies labeled 
with {\tt Code 1}, which are detected in both surveys. 
We define the relative error of the flux for each galaxy 
\citep{2025ApJS..279...32Y}     
\begin{equation}
\delta S_{21}=\frac{S_{21,F}-S_{21,A}}{\sqrt{S_{21,F}S_{21,A}}},
\end{equation}
where $S_{21,F}$ and $S_{21,A}$ denote the flux obtained in FATHOMER and ALFALFA, respectively. 
In \reffg{fig:hiflux}, we show the comparison of the flux between FATHOMER and ALFALFA,
where the direct comparison of the flux measurements is shown in the top panel,
and the histogram of the relative errors is shown in the bottom panel.
In the top panel, the diagonal dashed line represents the equality
of the two flux measurements.
It is shown that the galaxy flux obtained in FATHOMER is consistent with that obtained
in ALFALFA. According to the histogram statistical analysis shown in the 
bottom panel of \reffg{fig:hiflux}, the mean absolute relative difference is 
$\langle|\delta S_{21}|\rangle \times 100\% \sim 17.7\%$. 

\subsection{The catalog completeness}\label{sec:comp}

The completeness of the FATHOMER survey is defined as the fraction of \hi galaxies detected by FATHOMER at a given integrated flux density and within the survey area, which is the same as the definition of the completeness in \cite{2011AJ....142..170H}.

The completeness is a function of flux $S_{21}$ and $W_{50}$. Given the limited sky coverage of the current FATHOMER survey and the small number of detected \hi\ galaxies,  $W_{50}$ is not further divided into multiple bins. We count the number $N$ in logarithmic intervals of flux density $S_{21}$ to measure the ${\rm d}N/{\rm d}\log S_{21}$ distribution. 
For a flux-limited sample with the uniform distribution of sources in the Euclidean space, the number counts $N$ should follow $N\propto S_{21}^{-3/2}$~\citep{1961MNRAS.122..389S,2011A&A...532A..49B,2011AJ....142..170H,2024arXiv241109903M}.      

\begin{figure}
\centering
\includegraphics[width=0.45\textwidth]{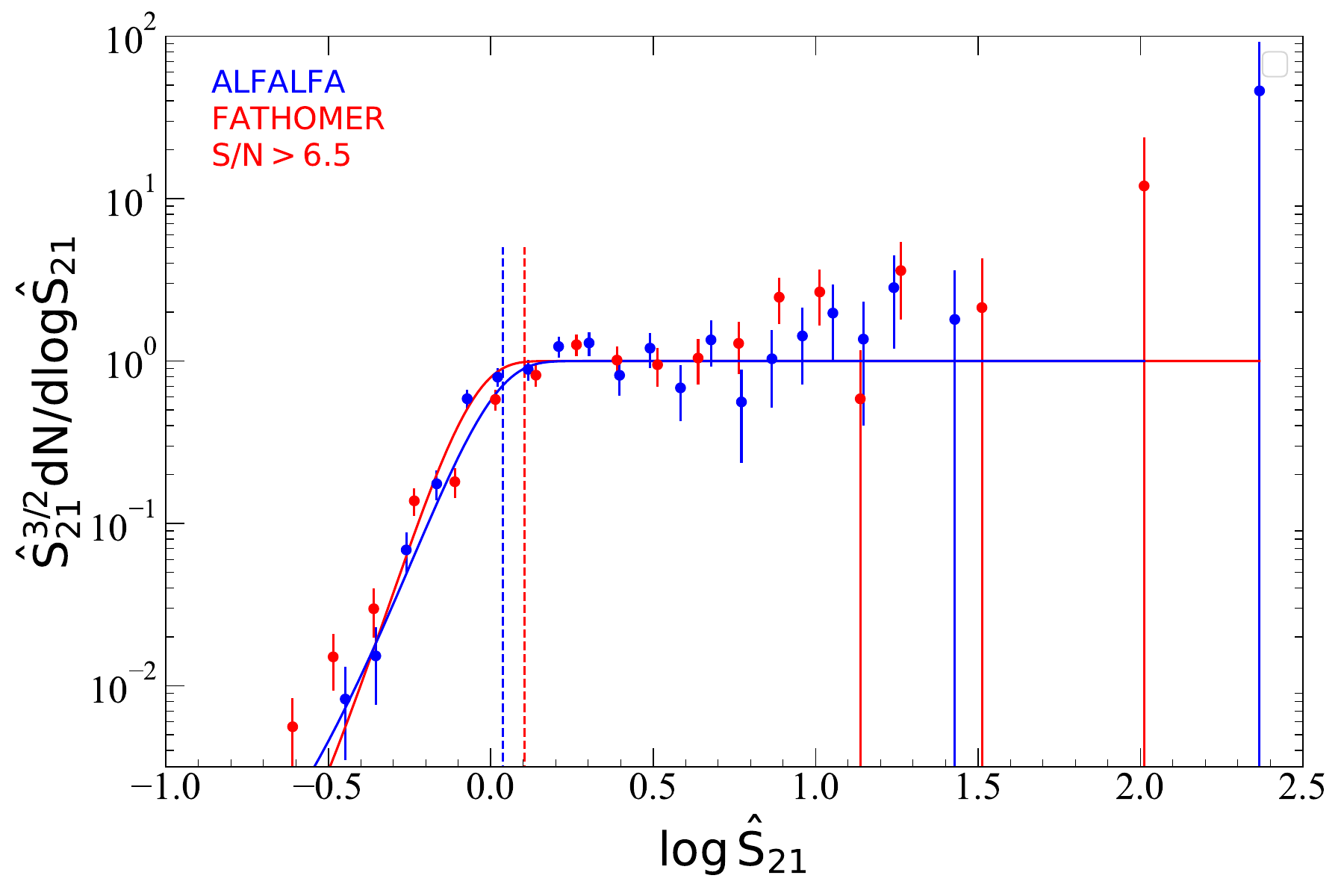}
\caption{Distribution of $\hat S_{21}^{3/2} {\rm d}N/{\rm d}\log \hat S_{21}$ versus $\log \hat S_{21}$ from the \hi galaxies with $\rm S/N \geq 6.5$, in the ALFALFA (blue) and FATHOMER (red) to evaluate the completeness.  In the FATHOMER, we can join the {\tt Code 1} and {\tt Code 2} candidates.
The vertical dashed lines represent the flux where the survey completeness is 90\%. The 90\% completeness is according to the fit of $\hat{S}_{21,90\%}$, and the correspond logarithmic values are 0.10 for FATHOMER and 0.034 for ALFALFA, respectively.
} 
\label{fig:dN/dlogS}
\end{figure}

We denote $\hat{S}_{21} = S_{\rm 21}/({\rm Jy} \cdot {\rm km\,s}^{-1})$, and in Figure~\ref{fig:dN/dlogS} we plot  $\hat{S}_{21}^{3/2} {\rm d}N/{\rm d}\log \hat S_{21}$ versus $\log \hat S_{21}$ for \hi galaxies labeled by {\tt Code 1}. 
We then fit the histogram using an error function $f_{\rm err}$, which is defined as  
\begin{equation}
    f_{\rm err}= A \times C(x) = A\times \frac{1}{2} \left[ 1+ {\rm erf}
    \left(\frac{x-x_{0}}{\sigma_x}\right) \right] ,
    \label{error function}
\end{equation}
where $C(x)$ is the completeness function, with $x=\log \hat S_{21}$, $A$ is the amplitude (or scale factor) of the function, $x_{0}$ is the location parameter, which shifts the curve along the x-axis, and $\sigma_x$ is a parameter controlling the steepness of the curve flank. We directly adopted the {\tt curve\_fit} function in {\tt SciPy}\citep{2020NatMe..17..261V} to perform the nonlinear curve fitting. After the best-fit curve is obtained for $f_{\rm err}$, we estimate the values of $\log \hat S_{21}$
at which $f_{\rm err}$ are 25\%, 50\%, and 90\% of its maximum.

The 25\% completeness threshold can be pinpointed as the integrated flux density limit; sources below this limit can be deemed false or unreliable signals. The 50\% completeness limit is the most relevant for the derivation of galaxy statistical distributions, such as the \hi mass function and the \hi width function. We can use it to analyze and study the HIMF when we obtain more \hi samples in the future \citep{2025arXiv250111872W}. The 90\% completeness indicates the position close to the knee of the curve for $\hat{S}_{21}^{3/2} {\rm d}N/{\rm d}\log \hat S_{21}$ versus $\log \hat S_{21}$. 

\reffg{fig:dN/dlogS} shows the measured $\hat{S}_{21}^{3/2} {\rm d}N/{\rm d}\log \hat S_{21}$ versus $\log \hat S_{21}$. 
We use the galaxies labeled as {\tt Code 1} and {\tt Code 2} in our galaxy catalog, 
and galaxies have SDSS spectroscopic counterparts in the ALFALFA catalog
(labeled as {\tt Code 1} in the ALFALFA galaxy catalog).
We abandoned the galaxies with S/N less than 6.5 and finally used
299 samples in FATHOMER and 350 samples in ALFALFA, respectively.
The solid red and blue lines represent the results for galaxies from  ALFALFA and FATHOMER, respectively. The vertical red and blue dashed lines in \reffg{fig:dN/dlogS} 
represent the $90\%$ completeness limit for the galaxies from ALFALFA and FATHOMER, respectively. It indicates that our survey can robustly reproduce the majority of high-confidence ALFALFA detections and slightly extend to fainter sources. As a result, the 90\% completeness turnover point derived from our source counts is very close to that of ALFALFA. This agreement also demonstrates that the reliability of our $\rm S/N > 6.5$ candidate selection.
Although the {\tt Code 2} galaxies in the FATHOMER catalogue are not detected by ALFALFA, these sources generally have lower flux densities and are limited in number. As a result, the sample completeness of FATHOMER and ALFALFA, as shown in \reffg{fig:dN/dlogS}, does not exhibit a significant difference.

As shown in the completeness analysis, the $90\%$ completeness limit for FATHOMER is 
$\log \hat{S}_{21} = 0.10$, slightly higher than the corresponding value of 
$\log \hat{S}_{21} = 0.034$ for ALFALFA. 
This difference is primarily attributed to the stronger radio frequency interference (RFI) present in the FATHOMER data, as discussed further in Section~\ref{sec:rfi}.

Interestingly, for $\log \hat{S}_{21}<-0.3$, the values of 
$\hat{S}_{21}^{3/2} , {\rm d}N/{\rm d} \log \hat{S}_{21}$ 
from FATHOMER exceed those from ALFALFA, indicating that FATHOMER detects 
more faint galaxies in the overlapping region despite the RFI limitations.

We also estimate the $50\%$ and $25\%$ completeness limits: the $50\%$ limits are
$\log \hat{S}_{21} = -0.025$ for FATHOMER and $-0.076$ for ALFALFA, while the 
$25\%$ limits are $-0.10$ and $-0.15$, respectively. The consistently higher completeness thresholds for FATHOMER reflect the impact of RFI on sensitivity, but they also highlight that FATHOMER still performs well in detecting faint sources, aided by FAST’s high sensitivity.

\subsection{Detection limit}

\begin{figure}
\centering
\includegraphics[width=0.47\textwidth]{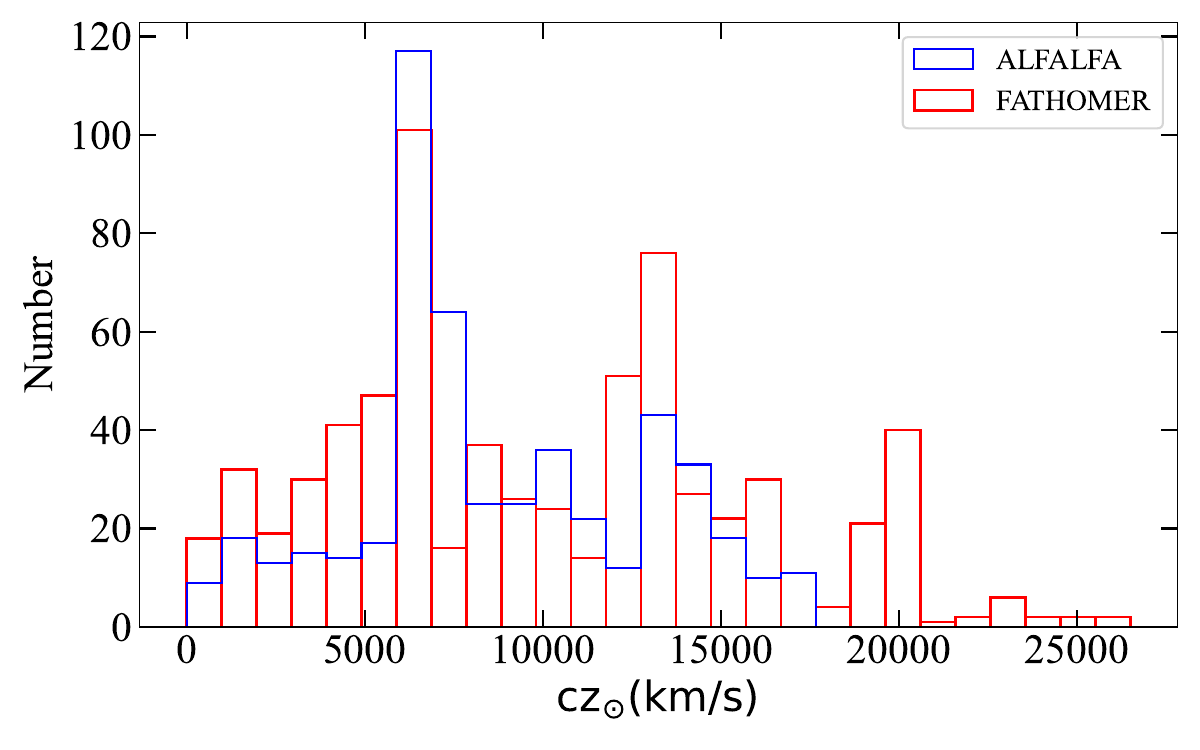}
\includegraphics[width=0.47\textwidth]{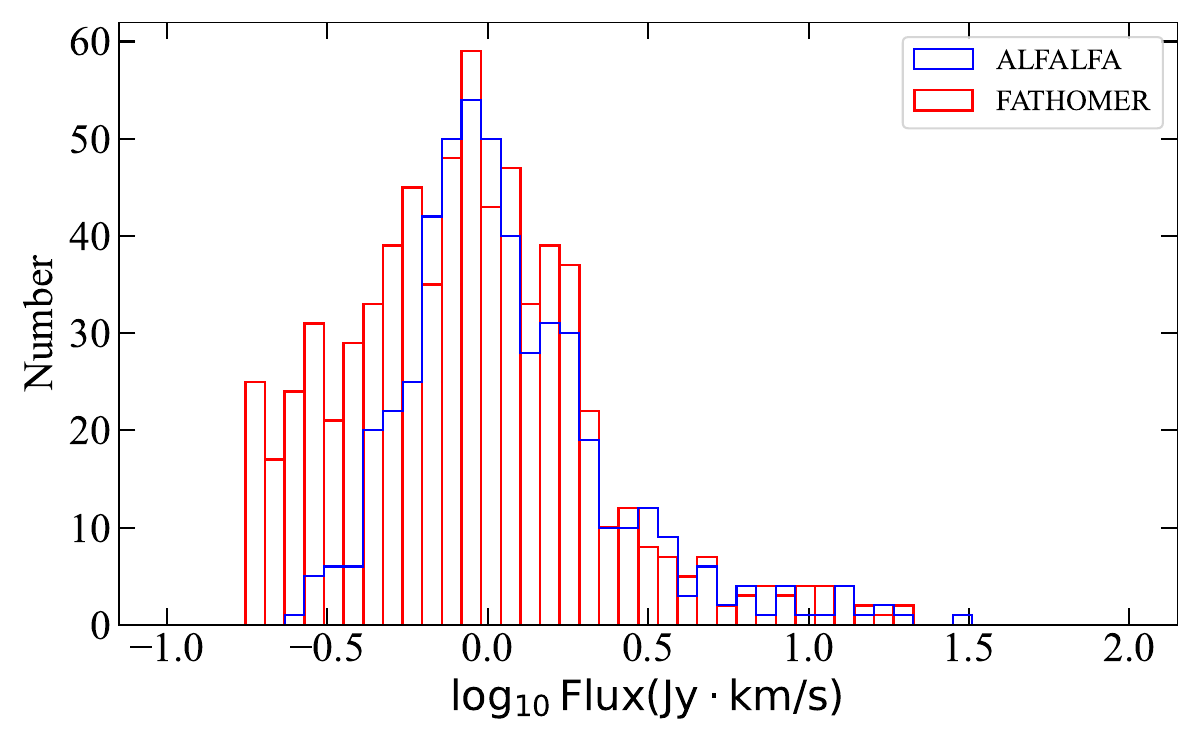}
\includegraphics[width=0.47\textwidth]{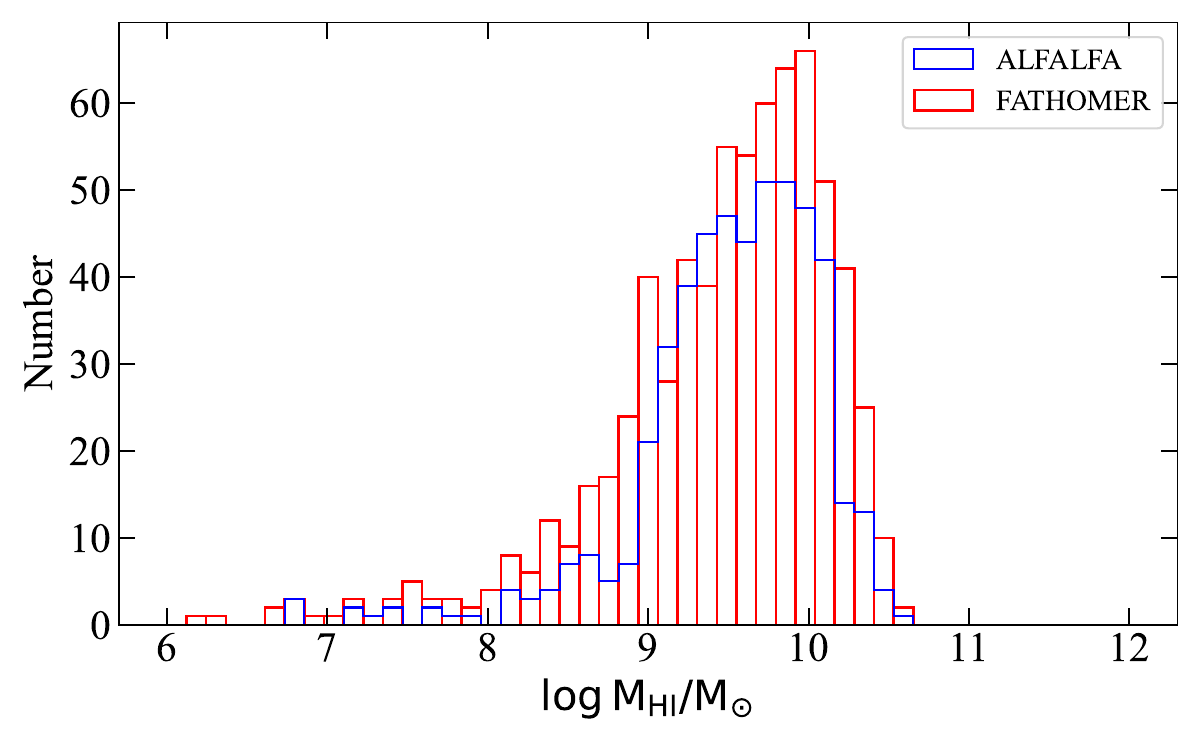}
\caption{{\it Top}: Histograms of the measured heliocentric velocity ($\rm cz_{\odot}$). 
{\it Middle}: Histograms of the measured \hi flux. 
{\it Bottom}: Histograms of the measured \hi mass. 
The results of the ALFALFA and the FATHOMER galaxy sample in the same sky area are shown in blue and red, respectively. 
}
\label{fig:cz_flux_hist}
\end{figure}

\begin{figure}
\centering
\includegraphics[width=0.5\textwidth]{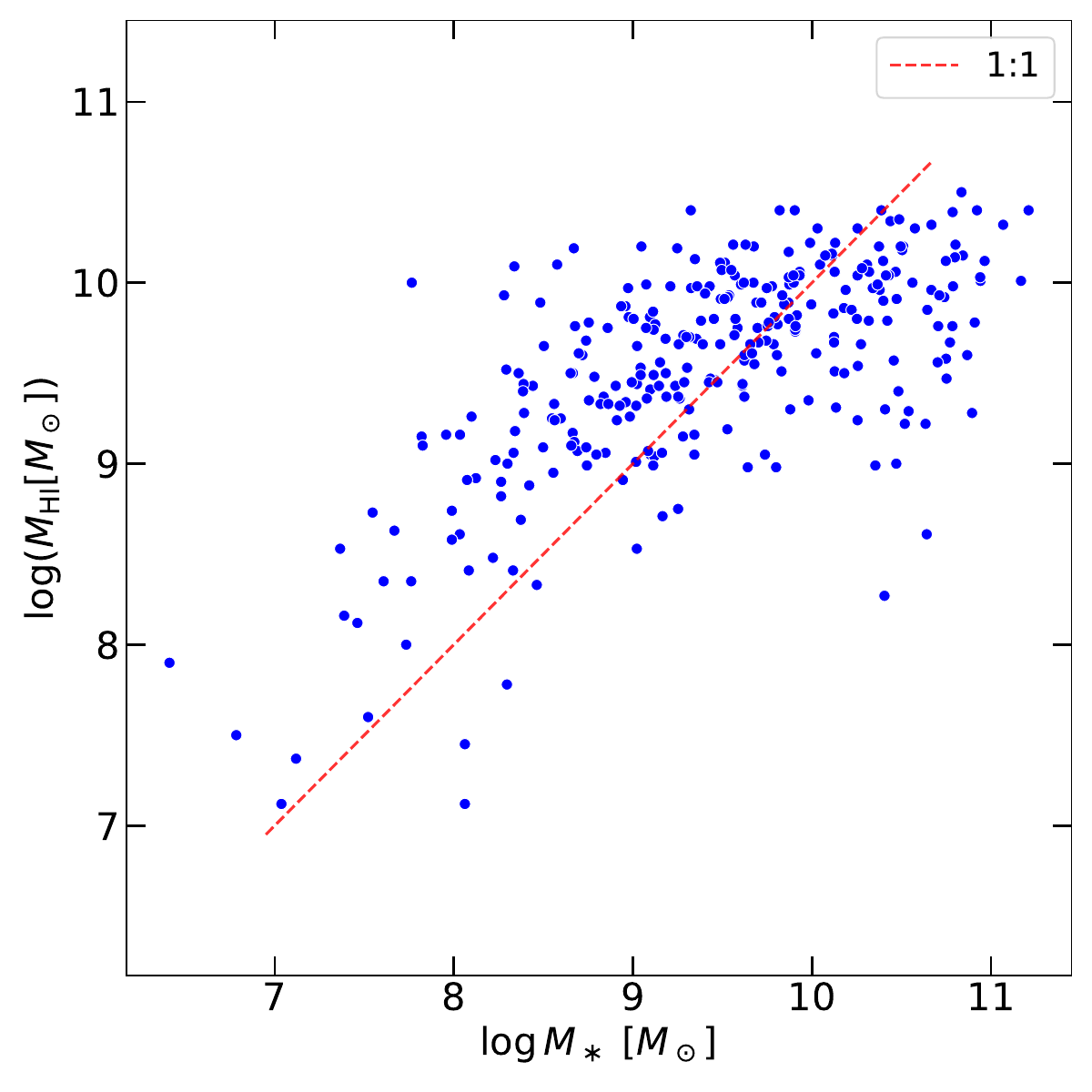}
\includegraphics[width=0.5\textwidth]{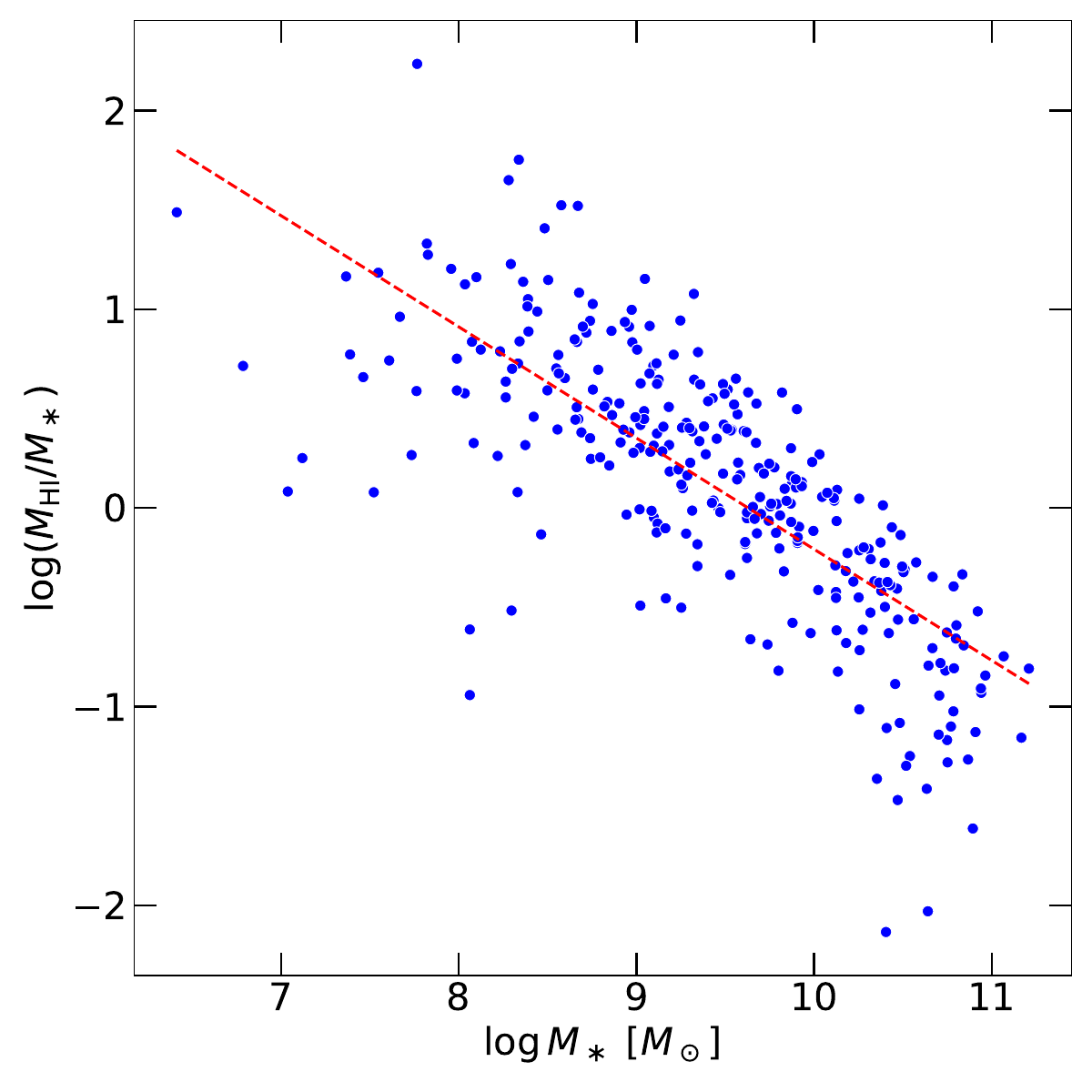}
\caption{Top: Correlations between the stellar mass $\rm M_{\ast}$ and \hi mass $\rm M_{\hi}$. Bottom:  \hi mass fraction as a function of stellar mass $\rm M_{\ast}$. The red line is the upper panel is 1:1 line and the red line in the lower panel is the line by fitting the data with $Y=AX+B$ function. The results shown here are for galaxies with code 1, 2 and 3.}
\label{fig:log_himass_to_stellarmass}
\end{figure}

The theoretical sensitivity of the telescope can be estimated by the thermal noise ~\citep{2008MNRAS.383..150D}:
\begin{equation}
    \sigma =\frac{2k_{\rm B}T_{\rm sys}}{\eta A_{\rm illu}\sqrt{N_{\rm pol}\Delta {t}\Delta{\nu }} },
    \label{sigma}
\end{equation}
where the $k_{\rm B}$ =1380 Jy $ \rm m^{2} \K^{-1}$ is the Boltzmann constant, 
$\eta = 0.6$ is the antenna efficiency, 
$A_{\rm illu} = 70700\,{\rm m^2}$ is the illuminated area,
and $N_{\rm pol} = 2$ is the number of polarizations.
We simply use the system temperature $T_{\rm sys} = 20\,{\rm K}$\citep{2020RAA....20...64J}, and 
the frequency resolution of $\Delta \nu = 30\,{\rm kHz}$. 
$\Delta t$ is the total integration time per pixel. 
The integration time per pixel is approximated as the time that a pixel takes to transit over a
beam width \citep{2020MNRAS.493.5854H},
\begin{equation}
t_{\rm p} = \frac{\theta_{\rm FWHM}}{\omega_{\rm e} \cos(\delta)},
\end{equation}
where $\theta_{\rm FWHM} \sim 2.9'$ is the FAST's beam width, 
$\omega_{\rm e} = 0.25\,{\rm arcmin\,s}^{-1}$, and $\delta\sim26^\circ$ is the
mean decl. of the target field.
During the drift-scan observation, the 19-feed array are rotated around the central feed
by $23.4^\circ$ to optimize the field coverage. The overlapping of different feeds 
increases the integration time by a factor of $f_{\rm OL} \sim 1.6$ on average
\citep{2025RAA....25e5008L}. 
In addition, we consider our current observation are carried out with overlapping
between different drift-scan strips, i.e. each source is been observed twice. 
Thus, the total integration time per pixel is estimated as,
\begin{equation}
\Delta t = 2 f_{\rm OL} \times t_{\rm p} \sim 41\,{\rm s}.
\end{equation}
The corresponding pixel thermal noise of the map is about $\sigma = 0.82\,{\rm mJy}$ and
the detection threshold of the frequency integrated \hi flux can be estimated by
\citep{2018ApJ...861...49H}:
\begin{equation}
      F_{\rm th}=S/N\times \sigma \times \sqrt{2W_{50}\Delta v } 
\label{fluxlimit function},
\end{equation}  
where we adopt $W_{50} = 200\,{\rm km\,s}^{-1}$, S/N = 5, and the spectral resolution 
$\Delta v = 6.3\,{\rm km\,s}^{-1}$. 
The corresponding detection threshold of the \hi galaxy is 
$F_{\rm th}\approx0.21{\rm Jy}\cdot{\rm km\,s}^{-1}$. In our detected galaxy sample, the minimum \hi flux is $0.21 {\rm Jy}\cdot{\rm km\,s}^{-1}$, which is consistent with the theoretical estimate. 
Compared to the ALFALFA \hi galaxy catalog, which has the thermal noise about $1.86\,{\rm mJy}$ 
and the spectrum  resolution of $10$ \kms \citep{2018ApJ...861...49H}, 
the detection limit of the \hi flux is $\sim 0.59\,{\rm Jy}\cdot{\rm km\,s}^{-1}$. 

With the full galaxy sample detected in FATHOMER, i.e., galaxies labeled with 
{\tt Code 1}, {\tt Code 2},{\tt Code 3}, and {\tt Code 4},
we show the histograms of the heliocentric velocities,
the \hi flux, and the \hi mass in the top, middle, and bottom panels of
\reffg{fig:cz_flux_hist}, respectively.
The results of this work and the ALFALFA \hi galaxy sample
\citep{2018ApJ...861...49H} of the same sky area are shown in blue and red, respectively. 
All galaxies in FATHOMER have $S/N\ge5$. 

In order to estimate the stellar mass of each \hi galaxy, we adopt the \texttt{lgm\_tot\_p50} parameter from the MPA--JHU DR8 catalog and $\rm sdss\_dr12.stellarmass\_granada$ as the total stellar mass. This parameter represents the median of the posterior distribution of stellar mass, derived from Bayesian SED fitting based on the SDSS total galaxy photometry and the stellar population synthesis library of \citet{2003MNRAS.344.1000B}. In Figure ~\ref{fig:log_himass_to_stellarmass}, we show the relation between the stellar mass $\rm M_{\ast}$ and \hi mass $\rm M_{\hi}$ (upper panel), and relation between stellar mass and \hi mass fraction (bottom panel). We find that galaxies with higher stellar mass also have more massive \hi mass. However, the ratio of \hi mass to stellar mass is smaller than one for more massive galaxies. This result is consistent with the findings of \cite{2013MNRAS.436...34C,2014yCat..74360034C,2018MNRAS.476..875C}. 

As shown in the top panel of \reffg{fig:cz_flux_hist}, the FATHOMER extends the redshift
range of detected \hi galaxies compared to ALFALFA. 
While both surveys cover similar populations at low redshift, 
FATHOMER identifies several galaxies with $cz_\odot$ exceeding $20,000\, {\rm km\,s}^{-1}$,
with the most distant reaching $\sim 26,000\, {\rm km\,s}^{-1}$. 
For context, ALFALFA employs the ALFA receiver, which operates in the L-band (1225–1525 MHz). However, to minimize vulnerability to RFI, the actual observing frequency range is restricted to 1335–1435 MHz. Frequencies below 1335 MHz are heavily contaminated. The strongest and most persistent RFI feature arises from the FAA radar at the San Juan airport, which is centered near 1350 MHz, making reliable detection of HI emission infeasible in these affected regions.
This suggests that FAST has a modest advantage in detecting more distant
\hi galaxies within the same sky region.

In terms of flux distribution, FATHOMER recovers more faint \hi sources, 
with detections extending to lower integrated fluxes
($\log S_{\hi}/{\rm Jy\cdot km\,s}^{-1} < –0.5$). 
The peak of the flux distribution is consistent between the two surveys, 
but FATHOMER shows a broader tail on the low-flux end, 
indicating slightly improved sensitivity to weak signals.

Similarly, the \hi mass distribution shows that FATHOMER includes more
low-mass galaxies, particularly those with $\log(M_{\hi}/M_\odot) < 8$,
which are largely underrepresented in the ALFALFA sample. 
This modest extension toward the low-mass end is important for improving 
our understanding of faint and gas-rich galaxy populations, 
especially in the context of dwarf galaxy studies.

\subsection{Impact of data masking}
\label{sec:rfi}

\begin{figure}
\centering
\includegraphics[width=0.5\textwidth]{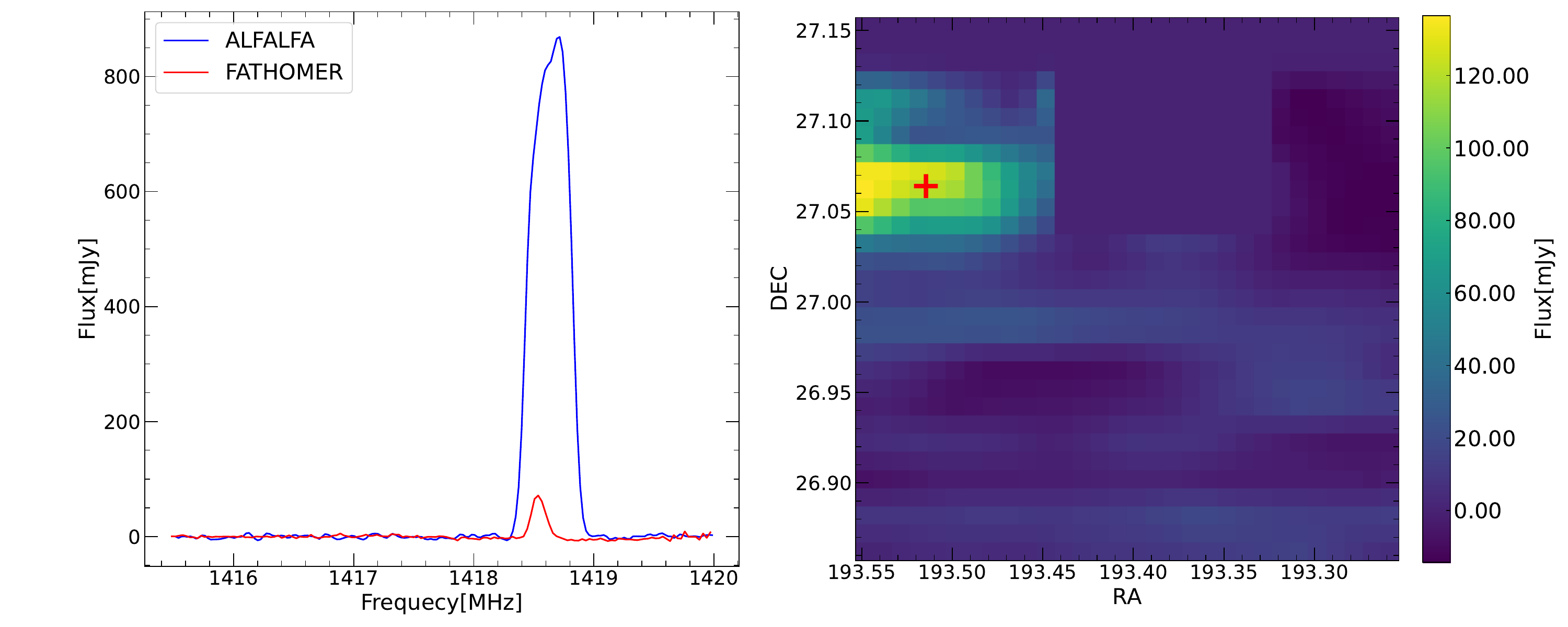}
\caption{Frequency spectra (left) and sky map (right) of the galaxy AGCNr 8024, for which the \hi line was not successfully recorded in FATHOMER. The blue and red lines represent the spectrum obtained by ALFALFA and FATHOMER, respectively. The right map is from FATHOMER, and the red cross labels the position of this galaxy. 
}
\label{fig:A8024}
\end{figure}

Within our sky coverage, ALFALFA has identified $491$ \hi galaxies with ${\rm S/N_{A}} \geq 3.5$. 
Among these, 350 galaxies have $\rm S/N_{A} \geq 6.5$, a threshold above which detections have been shown to be 
nearly $100\%$ reliable in the ALFALFA survey \citep{2011AJ....142..170H}. 
However, 160 of these 491 ALFALFA galaxies are not detected in the FATHOMER survey. 
These 160 non-detections can be further categorized as follows:
\begin{itemize}
\item 63 galaxies have $\rm S/N_{A} \geq 6.5$,
\item 56 galaxies have $5 \leq \rm S/N_{A} < 6.5$, and
\item 41 galaxies have $\rm S/N_{A} < 5$.
\end{itemize}

\begin{figure}
\centering
\includegraphics[width=0.45\textwidth]{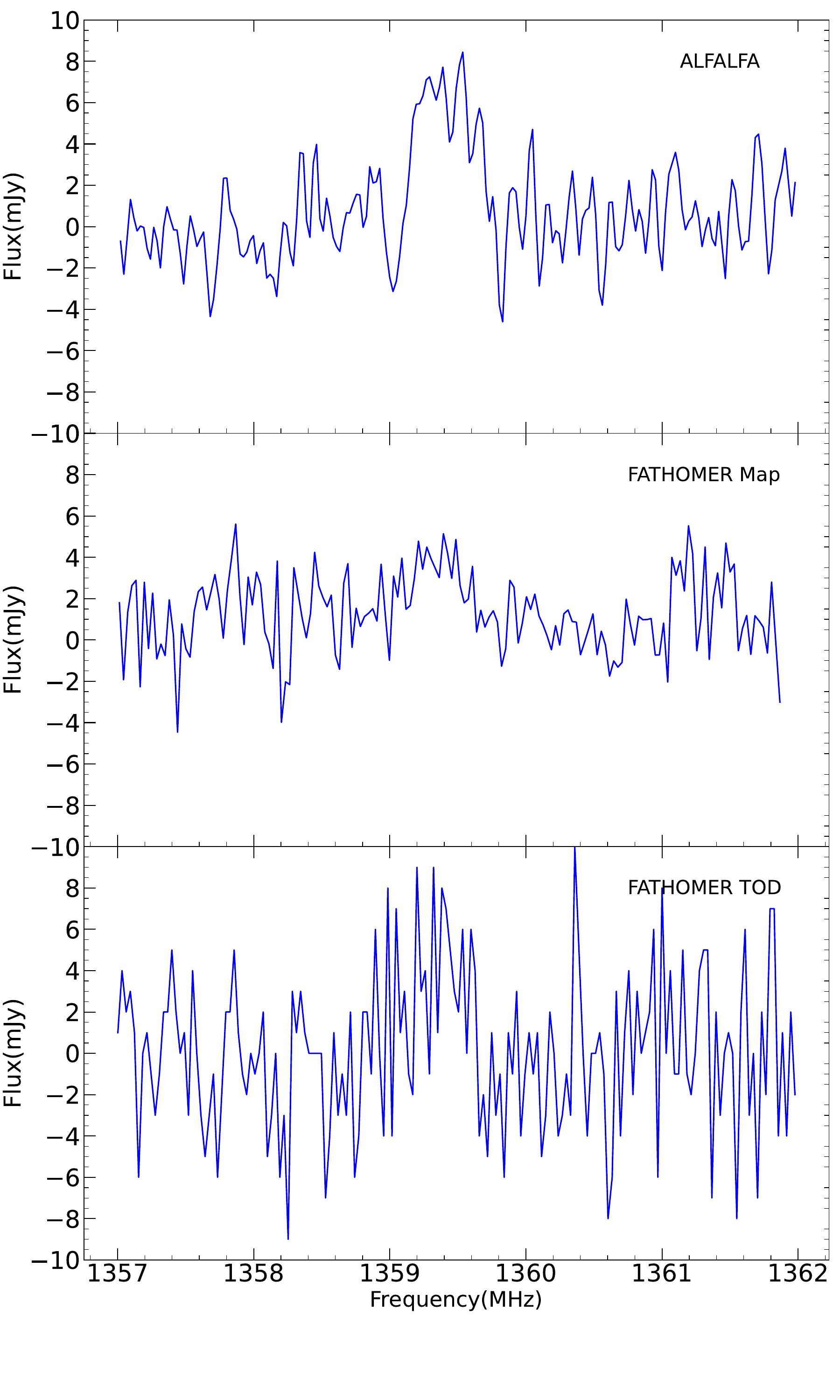}
\caption{Frequency spectra for galaxy AGCNr 194945 in ALFALFA (top), mapping data in FATHOMER (middle) and calibrated TOD (bottom).}
\label{fig:AGCNr194945}
\end{figure}

\subsubsection{Low-quality data masking}
Among the 63 galaxies with $\rm S/N_{A} \geq 6.5$ in ALFALFA, we found that there are 4 galaxies 
located at the edge of the FATHOMER observation field.
As an example, Figure~\ref{fig:A8024} shows a galaxy located at the edge of the FATHOMER field. 
The ALFALFA spectrum (red line) clearly indicates a high-S/N detection, confirming the galaxy’s authenticity. 
However, because only a portion of the galaxy was observed by the FAST beam (right panel), 
the corresponding signal in the FATHOMER spectrum is significantly weaker. 
As a result, it does not meet the $\rm S/N_{F} > 5$ threshold required for inclusion in {\tt Code 1}. 

In addition, another 9 galaxies, although they are not located at the edge of the FATHOMER field,
also have $\rm S/N_{F}$ below 5 in our FATHOMER data and are therefore excluded from the {\tt Code 1} catalog.
The relatively lower $\rm S/N$ in the FATHOMER survey than the ALFALFA survey may be due to the
varying noise level caused by the nonuniform survey coverage of the FATHOMER survey.
Such nonuniform survey coverage is primary due to the masking of the low quality TOD data.
In Figure~\ref{fig:AGCNr194945}, we show the spectra of galaxy AGCNr 194945. 
This galaxy has $\rm S/N_{A} = 7.5$ in the ALFALFA catalog and it is confirmed as
a prominent \hi galaxy with an optical counterpart. The spectrum obtained with ALFALFA
survey is shown in the top panel of Figure~\ref{fig:AGCNr194945}.
Such galaxy can be identified with the matched-filter method in both TOD and map of 
the FATHOMER survey. 
However, there is only one set of the available TOD. The rest are masked due to bad data quality. 
The spectrum of the available TOD is shown in the bottom panel of Figure~\ref{fig:AGCNr194945}.
Because of the lack of efficient observation, the $\rm S/N_{F}$ of this galaxy 
obtained from the FATHOMER map is $4.6$, which is below the detection threshold value ($\rm S/N=5$).

\subsubsection{RFI masking}

\begin{figure}
\centering
\includegraphics[width=0.49\textwidth]{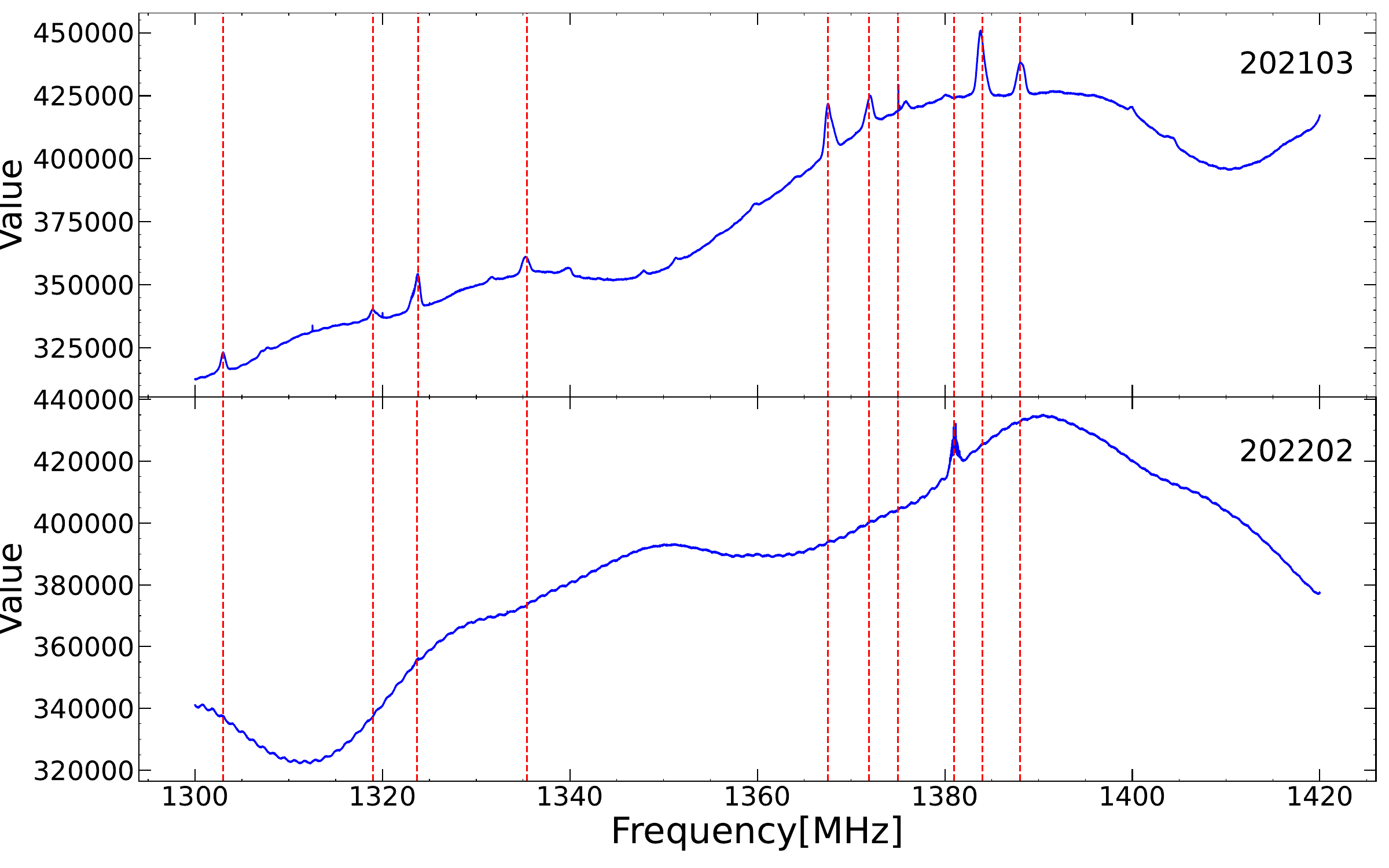}
\caption{Frequency spe
ctrum (polarization XX) from the central beam (blue line) in 2021 (upper panel) and 2022 (lower panel) observations.  The vertical red lines corresponds the frequencies at 1303.91MHz, 1320.11MHz, 1324.20MHz, 1336.40MHz, 1368.60MHz, 1372.49MHz, 1375.02MHz, 1381.05MHz, 1388.13MHz.}
\label{fig:rfi}
\end{figure}

\begin{figure}
\centering
\includegraphics[width=0.5\textwidth]{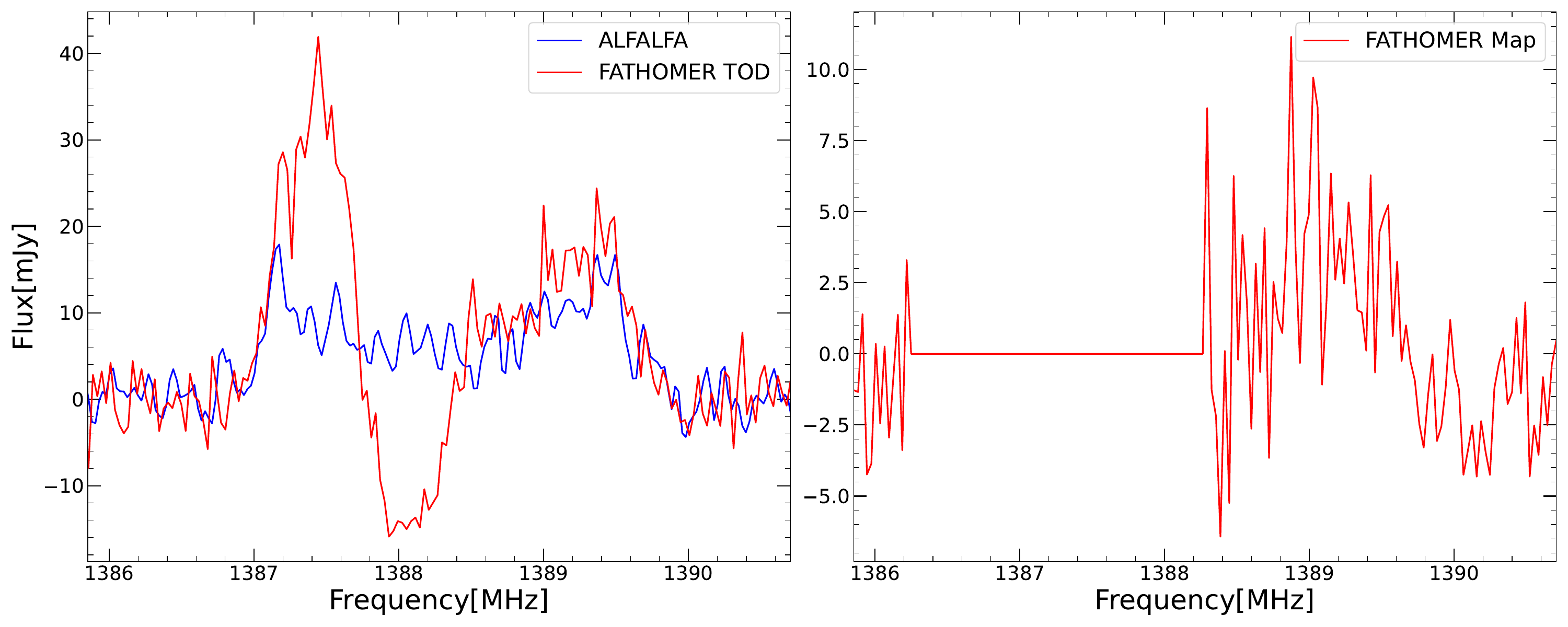}
\caption{Frequency spectra of the galaxy AGCNr 7419.
In the left panel, the red and blue lines represent the spectral from ALFALFA and calibrated TOD in FATHOMER, respectively.
In the right panel, the spectrum is from the mapping data in FATHOMER.}
\label{fig:A7419}
\end{figure}

\begin{figure}
\centering
\includegraphics[width=0.5\textwidth]{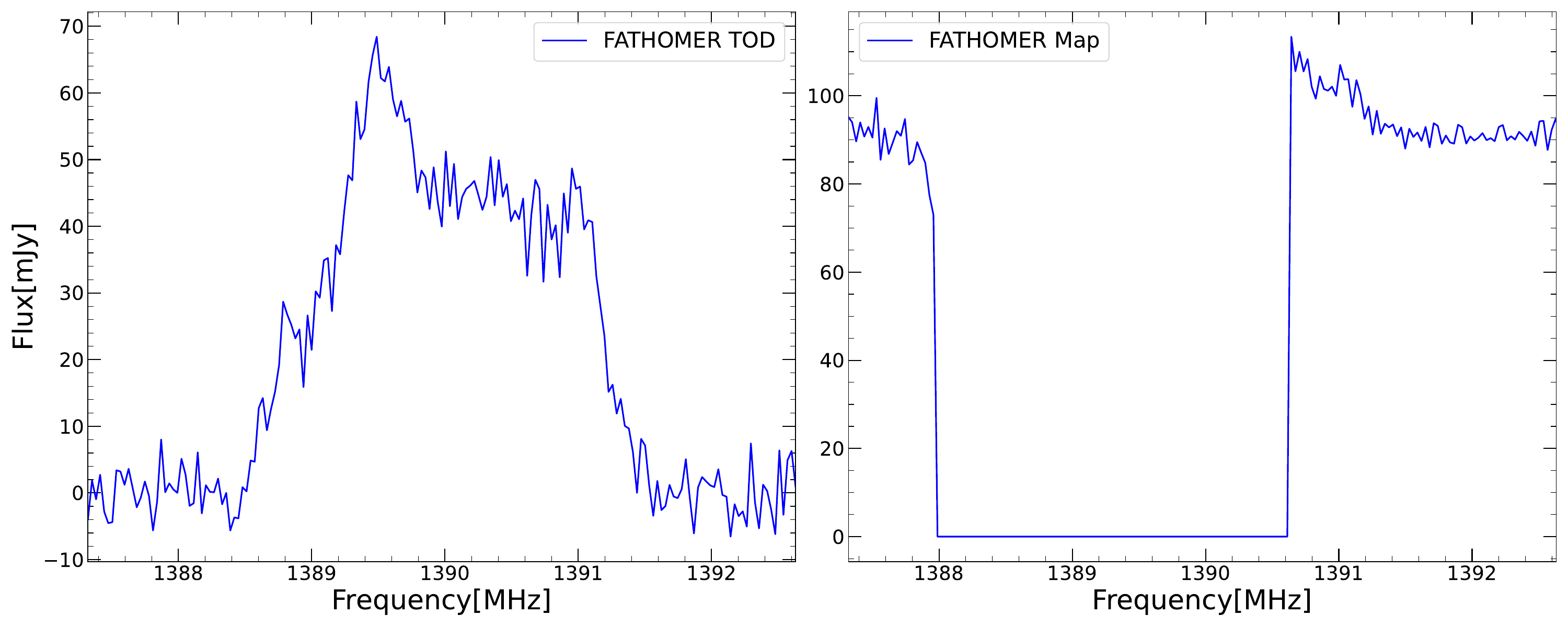}
\caption{Frequency spectrum of galaxy AGCNr 6190 in calibrated TOD  (left) and the mapping data (right). }
\label{fig:gal6190}
\end{figure}

The RFI contamination remains a major challenge in \hi galaxy detection. 
Among the 63 ALFALFA galaxies with $\rm S/N_{A} \geq 6.5$ that were not detected in FATHOMER, 
45 are significantly affected by RFI contamination. 
In addition, another 5 galaxies, despite having strong signals, were mistakenly labeled as RFI
and consequently masked during the mapping process.

Although a five-kilometer electromagnetic quiet zone has been established around the FAST site, 
strong RFI sources persist. Notably, the global positioning system (GPS) satellites 
near 1381.01 MHz have been identified as prominent RFI contributors 
\citep{2020RAA....20...75Z, 2021RAA....21...18W}. 
Even more problematic, some of the strongest RFIs were found to originate from the 
power supply of three compressors housed in the FAST feed cabin, which are used to 
cool the receiver system \citep{2025RAA....25a5011X}.

As illustrated in the upper panel of Figure~\ref{fig:rfi}, a series of RFI features appear
in the frequency spectrum from March 2021. Most of these features disappear in later data
from February 2022, except for a persistent signal near 1381.05 MHz. 
The interference from the compressors was successfully mitigated in August 2021 when 
shielding was applied to the compressor units by the FAST engineering team. 
However, our observations were conducted in March 2021, prior to this mitigation, 
so galaxies with frequencies overlapping with these RFIs were masked during the RFI flagging process.

In addition to the compressor-related RFIs, several other persistent RFI signals
have been identified in the 2022 data, likely originating from satellite transmissions.
These include:1317.00 MHZ,
1323.70 MHz, 1333.23 MHz, 1336.21 MHz, 1348.10 MHz, 1353.60 MHz, 1357.66 MHz, 1360.00 MHz, 1367.87 MHz, 1374.11 MHz, 1381.05 MHz, 1384.51 MHz, 1392.69 MHz, 1401.97 MHz, 1404.53 MHz, 1405.75 MHz, 1407.27 MHz, 1415.90 MHz, and 1417.17 MHz.
Among them, 1381.05 MHz corresponds to the GPS L3 band \citep{2021RAA....21...18W}, 
and the band from 1400 to 1427 MHz overlaps with remote sensing satellite transmissions. 
These frequencies require particular caution when selecting \hi galaxy candidates.

Figure~\ref{fig:A7419} shows the frequency spectrum of the galaxy AGCNr 7419. 
Two strong RFI features at 1387.5 MHz and 1389 MHz fall on either side of the galaxy, 
and the automated pipeline flags both the RFIs and the galaxy as contaminated. 
As a result, the galaxy is not recovered in the final map. 
In the calibrated time-ordered data (TOD), the strong RFI makes it impossible to 
extract accurate galaxy parameters. Such galaxies are categorized as 
{\tt Code 5} in our classification system.

In contrast, Figure~\ref{fig:gal6190} presents a case where a galaxy was misidentified as RFI. 
The left panel shows the spectrum from the calibrated TOD, 
while the right panel shows the spectrum from the final map. 
In both FATHOMER and ALFALFA data, this galaxy exhibits a high S/N ($>30$). 
In fact, all five galaxies that were mistakenly labeled as RFI in FATHOMER have
$\rm S/N > 15$, indicating that they are real sources wrongly flagged due to 
overly conservative RFI masking.
In our data processing pipeline, we attempt to distinguish real galaxy signals from RFI by leveraging the fact that real astrophysical sources are typically not detected simultaneously by multiple beams. To avoid misclassifying genuine signals as RFI, we average the signals from different beams at the same timestamp \citepalias{Li2023}. However, for nearby galaxies in the local Universe, this strategy becomes ineffective: their fluxes are often too strong, and their angular sizes can exceed the beam separation, making them visible in multiple beams simultaneously. As a result, our RFI mitigation method may inadvertently suppress these real signals.

\subsection{New low-redshift HI candidates in FATHOMER}
We investigate the intriguing subset of sources that are detected by FATHOMER but not by ALFALFA, despite lying within the redshift and flux sensitivity limits of ALFALFA’s survey. It is known that ALFALFA can only detect the galaxies with redshit smaller than 0.06 due to the RFI from the San Juan airport radar ~\citep{2015A&ARv..24....1G}. Therefore, we only consider the galaxies with redshfit smaller than 0.06 here. We found that in our survey area, there are 5 {\tt Code 2} and 216 {\tt Code 3} galaxies with $\rm cz_{\odot}$ $\leq$ 18,000 that were not detected by ALFALFA. 

Using Equation~\ref{fluxlimit function}, we estimated the $\rm S/N_{A}$ of the ALFALFA detections by substituting the ALFALFA detection limit (1.86 mJy), the thermal noise $\sigma$, the corresponding candidate’s W50, and the ALFALFA frequency resolution of 10 ${\rm km\,s}^{-1}$.
We find that 104 candidates fall below the ALFALFA detection limit, 
while 117 lie above the nominal ALFALFA sensitivity limit and are thus expected to be detectable. 
Among these 117 candidates, 54 have the $\rm S/N_A$ between 5 and 10, 
while 63 have ${\rm S/N_A}\geq 10$. A comparable situation was noted in the FASHI catalog analysis. In the overlapping sky region ($\rm 10 h < R.A. < 15 h$, $30^{\circ} < \rm decl. < 36^{\circ}$), FASHI detected 848 sources that were not seen by ALFALFA at $\rm cz_{\odot}$ $\leq$ 18,000. Among these 848 candidates, 724 have the $\rm S/N_A$ between 5 and 10, 
while 124 have ${\rm S/N_A}\geq 10$.

The robust $\rm S/N_A$ of our candidates suggest they are genuine HI sources. 
The fact that some remain undetected in ALFALFA can be attributed to several practical factors, including: ($\mathrm{i}$) the theoretical detection limit only provides an idealized threshold, as localized fluctuations in the thermal noise ($\sigma$) can significantly degrade the actual sensitivity within specific frequency bands; ($\mathrm{ii}$) 
environmental RFI and baseline instabilities may lead to localized threshold variations, reducing detection efficiency;
local variations in noise, baseline instabilities, and RFI contamination may reduce the effective sensitivity;
($\mathrm{iii}$) The matched-filtering reliability analysis, the completeness drops significantly at $\rm S/N_A<10$ levels.
We will process additional observation data in the future to ensure the reliability of these candidates.

\subsection{Cross-check detections between searching methods}

We have employed two methods to search for \hi galaxies: matched-filtering and \sofiaii. 
Comparing the results from these two approaches helps assess the reliability of
\hi candidates, particularly those with low S/N.

To assess the reliability of the two source-finding algorithms, we re-ran the \hi candidate search on the data cube after inverting its flux (multiplying by -1). The same source-finding procedure, including baseline subtraction, median filtering, and SNR thresholding, was applied to the negative cube. 
The inverted data cube analysis yielded 25 candidate sources. 
Compared to the total sample of 702 found with the original data cube, 
such non-zero detection in the inverted data cube indicates that $\sim 4.4\%$ detections in the original data cube
may be affected by noise fluctuations and RFI.

Table~\ref{table:catalog} presents the comparison between matched filtering and \sofiaii. 
For galaxies labeled as {\tt Code 1} and {\tt Code 2}, both methods yield identical detections.
These sources exhibit clear \hi profiles and stand out well above the noise, 
making their identification straightforward.

Among the more ambiguous {\tt Code 3} sources, 279 \hi candidates are cross-identified
by both methods. Notably, all galaxies detected by \sofiaii in this category are 
also detected by the matched-filtering method. 
Additionally, matched filtering identifies 26 extra {\tt Code 3} candidates that
are not picked up by \sofiaii. 
These additional detections tend to have low S/N values, ranging from 5.0 to 6.2.

We have tested the sensitivity of our results to the choice of smoothing parameters in both methods.
In \sofiaii, the initial maximum spectral kernel is set to 31, which corresponds to $\sim195\,\mathrm{km\,s^{-1}}$(the resolution of the spectrum is $6.3\,\mathrm{km\,s^{-1}}$), while the matched filtering extends to $\sim253\,\mathrm{km\,s^{-1}}$. 
We find that results of \hi galaxies obtained with a maximum spectral kernel of 40 are the same as those obtained with a kernel of 31.
It is easy to understand this fact since most galaxies in our catalog have line widths below $150\,\mathrm{km\,s^{-1}}$, which is well within the range covered by both methods. 
The pixel size of our maps is $1.72'$. The spatial direction is initially smoothed by 5 or 10 pixels. Given the telescope beam of $3'$, smoothing over 5--10 pixels ($15$--$30'$), being much larger than the beam, does not recover additional sources. We have checked that even when we added a 3-pixel parameter to search for candidates and applied candidate selection, no significant differences were found. 

Simulation results indicate that the matched-filtering method can reliably identify
galaxies with S/N above 5.5, achieving over $98\%$ confidence in the presence of Gaussian noise.
Even for broader detections with S/N between 4.5 and 5.5, the reliability remains high
at approximately $91\%$ \citep{2007AJ....133.2087S}. 
On the other hand, mock galaxy experiments demonstrate that SoFiA is capable of
producing a 100\% reliable output catalog across all S/N levels \citep{2021MNRAS.506.3962W}.

In summary, the matched-filtering approach is more sensitive and capable
of detecting fainter \hi\ candidates at lower S/N, whereas \sofiaii offers
higher reliability, especially for ensuring that detected sources are real.

\subsection{Potential Contamination of OH megamasers}
OH megamasers (OHMs) are luminous 18 cm molecular masers which are typically found in ultra luminous infrared galaxies and are usually taken as indicators of major galaxy mergers. The emission of OHMs occurs predominantly in the main lines at 1665 and 1667 MHz. 
The profiles of OHMs and \hi galaxies are similar. Therefore, it is difficult to distinguish them ~\citep[e.g.][]{2016MNRAS.459..220S,2021ApJ...911...38R}. 

Since \hi galaxies with code 3 and code 4 in our survey lack optical spectroscopy, these galaxies are likely to be contaminated by OHMs. In order to remove the possible OHMs from our candidates, we first cross-matched the galaxies with code 3 and code 4 with the known OHMs in our survey regions ~\citep{2002AJ....124..100D,2016MNRAS.459..220S,2018ApJ...861...49H,2024ApJ...971..131Z}. No known OHMs were found in our code 3 and code 4 samples.
We then followed the method described in  ~\cite{2016MNRAS.459..220S}, which uses infrared colors and magnitudes (see their Table 7), to identify potential OHMs. None of our \hi galaxies met the criteria for being OHMs. Therefore, the possibility that our \hi galaxies are contaminated by OHMs is negligible.

\subsection{Preliminary results of the \hi mass function}\label{sc:himf}
The \hi mass function  (HIMF) describes the number density of \hi galaxies at a given \hi mass, which not only reflects the total \hi mass in the universe, but also provides information about galaxy assembly histories.      
Since FAST  has the ability to detect weak signals, this provides us with the possibility to measure the HIMF at the low mass end.

\begin{figure}
\centering
\includegraphics[width=0.5\textwidth]{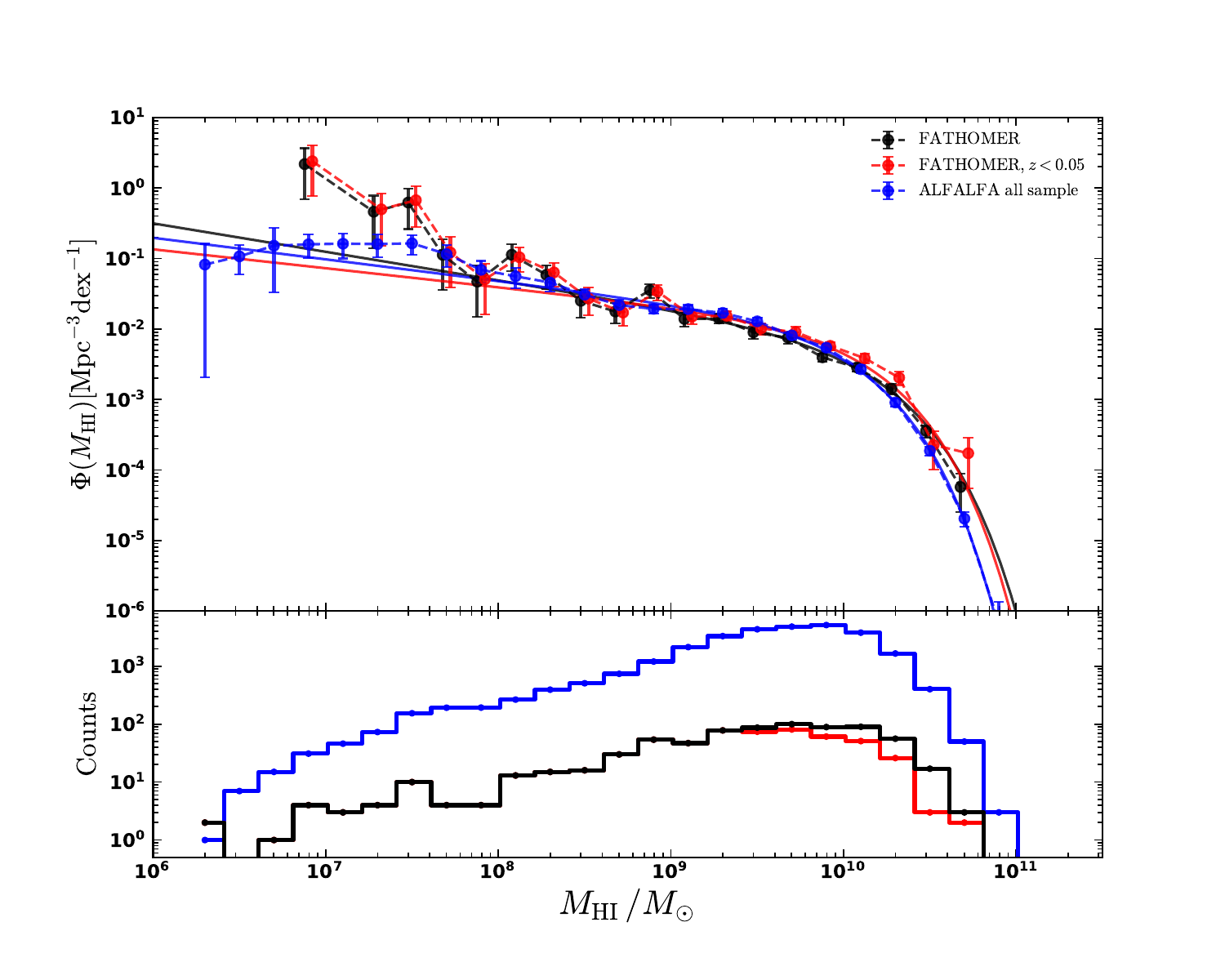}
\caption{{\it Top panel:} 
The \hi mass function estimated using galaxies of the ALFALFA survey is shown in blue, in the FATHOMER catalog with a redshift threshold of 0.05 is shown in red, and in the FATHOMER catalog without a redshift threshold is shown in black, respectively.
{\it Bottom panel:} shows the corresponding \hi mass histogram of the galaxies in the FATHOMER catalog with redshift threshold, without redshift threshold, and the ALFALFA catalog.}
\label{fig:himf}
\end{figure}

\begin{table}
\scriptsize
\begin{center}
\caption{The best-fit parameters for the \hi mass function.}\label{tab:himf_p}
\hspace*{-1.0cm}
\begin{tabular}{lccc} \hline\hline
Survey             &       ${\rm log}_{10}\Phi_*$             & $\log_{10}M_*/M_\odot$ & $\alpha$ \\\hline
FATHOMER           & $-2.51^{+0.09}_{-0.08}$ & $10.09^{+0.05}_{-0.05}$  & $-1.40^{+0.06}_{-0.07}$ \\ 
FATHOMER $z<0.05$  & $-2.31^{+0.10}_{-0.09}$ & $10.02^{+0.06}_{-0.06}$  & $-1.27^{+0.08}_{-0.10}$ \\ 
ALFALFA all sample & $-2.27^{+0.04}_{-0.04}$ & $ 9.93^{+0.02}_{-0.02}$  & $-1.31^{+0.03}_{-0.03}$ \\
\hline\hline
\end{tabular}
\end{center}
\end{table}


We adopt the modified $\sum1/V_{\rm max}$ method \citep{1968ApJ...151..393S, 10.1046/j.1365-8711.2001.04078.x, 2021MNRAS.501.4550X} 
to estimate the HIMF, incorporating corrections for survey completeness. The estimator is defined as, 
\begin{align}
\hat{\Phi}(M_{\hi}) = \frac{1}{{\rm d}\log{M_{\hi}}}\sum_{i=1}^{N_{\rm gal}} \frac{1/(1-f_{i,{\rm RFI}})}{V_{i,{\rm max}}},
\end{align}
where $N_{\rm gal}$ is the total number of galaxies in the \hi mass bin ${\rm d} \log M_{\hi}$,
$f_{i,{\rm RFI}}$ is the fraction of RFI-flagged pixels (as shown in \reffg{fig:rms})
at the redshift bin of the $i$-th galaxy. 
The factor $1-f_{i,{\rm RFI}}$ is adopted to compensate the 
fraction of missing galaxies due to the RFI-flagging.
$V_{i,{\rm max}}$ is the completeness-weighted maximum accessible volume for the $i$-th galaxy. 
In particular, $V_{i,{\rm max}}$ is given by \citep{2015MNRAS.452.3726H}
\begin{align}
V_{i,{\rm max}} = \sum_{j=1}^{j_{\rm max}}(\Delta V_j\times C_j),
\end{align}
where $\Delta V_j$ is the comoving survey volume of the $j$-th redshift interval, 
$C_j$ is the corresponding completeness factor. 
The upper limit $j_{\rm max}$ is set by gradually increasing the redshift of the galaxy
until its flux density drops below the survey detection threshold. 
This threshold is evaluated for each galaxy individually, 
using its \hi line width and the local systematic noise level via \refeq{fluxlimit function}.

Completeness is estimated using \refeq{error function}. 
The \hi galaxy catalog is not a flux-limited sample, but also \hi line width dependent,
as the galaxies of narrower \hi line widths are more readily detected at a given flux density
\citep{2005AJ....130.2598G,2011AJ....142..170H}. 
Thus, the FATHOMER catalog is divided into three line-width sub-samples, 
for which completeness is evaluated separately. 
The division is limited to three bins due to the modest sample size. 
In the HIMF calculation, only galaxies with completeness above 
$50\%$ are retained \citep{Jones_2018}, which excludes $\sim 40\%$ of the catalog.

The top panel of \reffg{fig:himf} shows the preliminary results of the HIMF 
estimated using all the galaxies in the FATHOMER catalog, and the bottom panel
shows the \hi mass distributions of our catalog.
The results using the galaxies corresponding to {\tt Code 1}, {\tt Code 2}, {\tt Code 3}, and {\tt Code 4} with a redshift threshold of 0.05 are shown in red. The results without the redshift threshold, referred to as FATHOMER, are shown in black. 
Meanwhile, we also apply our HIMF estimator to the \hi galaxy catalog with $\rm S/N>5$ in the ALFALFA survey \citep{2018ApJ...861...49H}, which is shown in blue.

To estimate the uncertainty of the HIMF, we adopt the jackknife resampling method. 
The full survey area is divided into $N_{\rm JK}=40$ approximately equal subfields. 
In each iteration, one subfield is omitted, and the HIMF $\phi_i$ is recalculated using the remaining subfields.
The variance across these jackknife samples provides an estimate of the statistical uncertainty,
accounting for sample variance and spatial inhomogeneity within the survey field,
\begin{equation}
\sigma^2_{\rm JK} = \frac{N_{\rm JK}-1}{N_{\rm JK}} \sum_{i=1}^{N_{\rm JK}} 
\left( \phi_i - \bar{\phi} \right)^2,
\end{equation}
where $\bar{\phi} = \frac{1}{N_{\rm JK}} \sum_{i=1}^{N_{\rm JK}} \phi_i$ is the mean value of each estimation.

We observe a pronounced excess in the HIMF at the low-mass end when using the FATHOMER catalog. 
This may arise from uncertainties in the completeness correction. 
However, owing to the limited sample size, only three line-width bins are employed, 
which likely introduces additional uncertainty in the completeness estimation. 
Future work will refine the completeness treatment to mitigate these effects.
As long as the uncertainty is relatively large at the low-mass end, such measurements have
limited weight in the following HIMF model fitting.

The solid curves in the top panel of \reffg{fig:himf}
represents the best-fit HIMF, which is modeled using the 
Schechter function \citep{1976ApJ...203..297S,2005MNRAS.359L..30Z},
\begin{align}
    \Phi (M_{\rm \hi})=\ln(10) \, \Phi_*\bigg(\frac{M_\hi}{M_*}\bigg)^{\alpha +1}
    e^{-\frac{M_\hi}{M_*}},
\end{align}
where $\Phi_*$ represents the normalization constant, $M_*$ characterizes the 'knee' mass of the HIMF,
and $\alpha + 1$ is the faint-end slope.
We show the best-fit parameters for the \hi mass function in \reftb{tab:himf_p}.

In this work, we adopt a cosmology with $h = 0.67$ when estimating the HIMF. 
The characteristic knee mass $M_*$ derived from the FATHOMER catalog is $\log_{10} M_* = 10.09$. 
For comparison, converting previous survey results to the same $h = 0.67$ cosmology yields $\log_{10} M_* = 9.90$ for HIPASS \citep{2005MNRAS.359L..30Z} and $\log_{10} M_* = 9.98$ for ALFALFA \citep{2018MNRAS.477....2J}. 
Thus, the FATHOMER value is about 0.11 dex higher than that of ALFALFA and 0.19 dex higher than that of HIPASS. 

This modest offset is well within the variations reported across different \hi surveys. 
The higher sensitivity and larger volume within the same sky region probed by FAST make FATHOMER more effective at detecting rare, massive \hi galaxies, naturally shifting the fitted $M_*$ upward compared to shallower surveys like HIPASS. 
Moreover, survey-to-survey differences in sky coverage and environment (cosmic variance) are known to introduce variations of order 0.1--0.2 dex in $M_*$. 
Therefore, the slightly larger $M_*$ found here is consistent with expectations and lies comfortably within the scatter established by existing measurements. 
We also note that the faint-end slope $\alpha$ measured in HIPASS ($\alpha \approx -1.37$) and ALFALFA ($\alpha \approx -1.25$) are broadly consistent, suggesting that the main difference among surveys arises in the determination of $M_*$ rather than in the low-mass slope.

\section{Summary and discussion} \label{sec:summary}
In this work, we present the results of a blind \hi galaxy survey conducted
over a continuous $60\,\deg^2$ region using the FAST telescope, 
as part of the FATHOMER project. 
We employed two complementary source-finding algorithms, i.e. the matched-filter approach and 
\sofiaii, to identify \hi galaxies from the observed map. 
These two methods show strong agreement, with all high S/N detections 
(labeled as {\tt Code 1} and {\tt Code 2}) being consistently identified. 
The matched-filter, being more sensitive, detects 26 additional low S/N candidates 
({\tt Code 3}), while \sofiaii provides higher reliability, especially at the faint end.

A total of 702 \hi sources were identified with {\rm S/N $\geq$ 5}, 
of which 331 are confirmed by ALFALFA, 
9 have spectroscopic counterparts in SDSS (new confirmations),
285 are matched to SDSS photometric objects and DESI, and
77 have no optical counterparts, potentially representing dwarf or dark galaxy candidates.

Despite the early FAST data being significantly affected by RFI, 
particularly from instrumental sources such as the feed cabin compressor, 
the detection rate reached 12.0 sources per degree square, 
nearly double that of ALFALFA in the same sky region. 
The fluxes of overlapping detections between FATHOMER and ALFALFA agree within 
$17.7\%$, validating our calibration and map-making pipeline (fpipe).

The current detection limit reaches down to HI mass of $10^{6.2}\, M_{\odot}$ 
and redshifts up to $z \sim 0.087$. Completeness and flux comparisons show that 
FATHOMER detects more faint galaxies than ALFALFA, though its completeness 
is limited by RFI. With hardware improvements and better RFI suppression since 2021,
future FAST observations are expected to yield higher completeness and deeper sensitivity.

We also present a preliminary estimate of the HI mass function (HIMF). 
The FATHOMER data yield a higher characteristic mass ($M_*$) and a steeper low-mass slope 
than the ALFALFA HIMF, likely due to 
FATHOMER’s ability to probe slightly higher redshifts.
Although these HIMF measurements are subject to uncertainties from incompleteness and RFI,
they demonstrate the potential of FAST-based surveys for cosmological and galaxy evolution studies.

This work demonstrates FAST’s strong potential for low-redshift \hi surveys. 
With ongoing and future observations covering larger sky areas and reduced RFI
contamination, FAST will enable the detection of fainter galaxies, 
including dark or optically elusive systems, 
and support detailed studies of the low-mass galaxy population and cosmic \hi distribution.

\section*{Acknowledgments} \label{sec:acknowledgements}

The authors thank Chen Xu, Qingze Chen, Yingjie Jing, Bo Zhang, Zheng Zheng, Yinghui Zheng and Chuanpeng Zhang for helpful discussion. 
This work is supported by the National SKA Program of China (Nos. 2022SKA0110100, 2022SKA0110101, 2022SKA0110200, 2022SKA0110203), the NSFC International (Regional) Cooperation and Exchange Project (No. 12361141814), the NSFC Innovation Group Project (No.12421003), the Specialized Research Fund for State Key Laboratory of Radio Astronomy and Technology, and the National Astronomical Observatories, Chinese Academy of Science (No. E5ZB0901). This work is also supported by the China Manned Space Program with grant no. CMS-CSST-2025-A02. YL acknowledges the support of the National Natural Science Foundation of China (No. 12473091) and
the Fundamental Research Funds for the Central Universities (No. N2405008).

This work made use of the data from FAST (Five-hundred-meter Aperture Spherical radio Telescope, \url{https://cstr.cn/31116.02.FAST}). FAST is a Chinese national mega-science facility, operated by National Astronomical Observatories, Chinese Academy of Sciences.

\section{DATA AVAILABILITY}
The result data underlying this article is available on Zenodo:https://doi.org/10.5281/zenodo.18782636

\bibliographystyle{aasjournal}
\bibliography{main}



\end{document}